\documentclass[oupdraft]{bio}
\usepackage{url} 
\usepackage{algorithm}
\usepackage{algorithmic}
\usepackage{mathrsfs}
\usepackage[margin=1in]{geometry}
\usepackage{graphicx}
\usepackage{amsmath, amssymb}
\usepackage{bbold}
\usepackage[colorlinks=true,allcolors=blue]{hyperref}
\usepackage{mathbbol}
\usepackage{float}
\usepackage{xcolor}
\usepackage{bm} 
\usepackage{setspace}
\usepackage{multirow}
\usepackage{lscape}
\usepackage[figuresright]{rotating}
\usepackage{adjustbox}
\usepackage{booktabs}
\usepackage{ragged2e}
\usepackage{subcaption}
\usepackage{float}
\usepackage{amsthm}
\usepackage{mathbbol}
\usepackage{tikz}
\usepackage{fix-cm}

\newcommand{\bpm}{\begin{pmatrix}}
\newcommand{\epm}{\end{pmatrix}}

\theoremstyle{definition}

\def\*#1{\mathbf{#1}}
\def\^#1{\mathbb{#1}}  
\def\+#1{\amsmathbb{#1}}
\def\##1{\mathbb{#1}}
\DeclareSymbolFontAlphabet{\amsmathbb}{AMSb}%

\begin{document}
\title{Penalized Cox Regression with Graph-Structured Distribution-Valued Predictors}

\author{Rahul Ghosal$^{1,\dagger,\ast}$, Dhrubajyoti Ghosh$^{2,\dagger}$, Yuanzhen Yue$^{1}$, and Arkaprava Roy$^{3}$ \\[4pt]
\textit{$^{1}$ Department of Epidemiology and Biostatistics, University of South Carolina \\
$^{2}$ Department of Data Science and Analytics, Kennesaw State University \\
$^{3}$ Department of Biostatistics, University of Florida}
\\[2pt]
{rghosal@mailbox.sc.edu}}

\markboth%
{R. Ghosal and others}
{GFDCM}

\maketitle
\footnotetext{To whom correspondence should be addressed. $^{\dagger}$Co-first authors.}

\begin{abstract}
{We propose the Graph-Fused Distributional Cox Model (GFDCM), a penalized survival analysis framework for distribution-valued predictors organized on a known graph. Modern neuroimaging studies collect high-dimensional voxel-level measurements across multiple regions of interest (ROIs) with rich spatial dependency structures. Rather than reducing these to scalar summaries, GFDCM models the full distribution of voxel intensities within each ROI, retaining region-specific interpretability while exploiting spatial relationships among adjacent ROIs. The distributional associations are modeled using penalized splines. To achieve simultaneous predictor selection and graph-based fusion, we introduce a structured Graph-Fused Group LASSO (GFGL) penalty that combines a group-sparsity term for discarding irrelevant predictors with a fusion term that enforces coefficient functions for graph-adjacent predictors to be identical. A flexible alternating direction method of multipliers (ADMM) algorithm is employed to optimize the penalized partial likelihood. Numerical simulations illustrate satisfactory selection, fusion, and estimation performance of GFDCM compared with unpenalized functional Cox regression and selection-only alternatives. Finally, we apply GFDCM to baseline T1-weighted MRI data collected from 308 patients in the Alzheimer’s Disease Neuroimaging Initiative (ADNI-1) to identify brain regions associated with AD conversion. The results provide clinically meaningful insights and exhibit superior predictive performance compared to traditional ROI-level summary-based approaches.}
{Distributional Cox Model; Graph Fused Group LASSO, Distributional Data Analysis; ADMM; Variable Selection; Alzheimer’s Disease; Neuroimaging; ADNI.}
\end{abstract}

\section{Introduction}
\label{sec:intro4}

Modern biomedical research increasingly relies on large-scale data sources that record biological variation across multiple spatial and temporal scales. In neuroimaging studies, each participant's brain is often partitioned
into anatomically defined regions of interest (ROIs), and within each
region, thousands of voxel-level intensity measurements capture tissue
density, microstructural integrity, and pathological burden
\citep{mueller2005ways, jack2010hypothetical, weiner2017alzheimer}.
These regional measurements are inherently high-dimensional and
heterogeneous.
Conventional analyses typically summarize each ROI by its mean
cortical thickness or average intensity
\citep{zhu2017novel, williams202112, ahmadzadeh2023neuroimaging,
ghosh2025ensemble}; however, such reductions may obscure clinically
meaningful variation contained in the spread, asymmetry, and tail
behavior of the underlying subject-specific intensity distribution.

For example, clinical and imaging evidence suggest that early neurodegenerative processes may manifest in specific portions of the distribution before shifts in average intensity become detectable. Two individuals may exhibit similar mean hippocampal intensity yet differ substantially in the lower quantiles of their distributions, reflecting early neuronal loss or microstructural disruption prior to visible atrophy \citep{sabuncu2011dynamics,jack2018nia}. In large neuroimaging studies, such as the Alzheimer’s Disease Neuroimaging Initiative (ADNI) \citep{petersen2010alzheimer,yu2022mapping}, practitioners are often interested in predicting time to conversion from mild cognitive impairment to Alzheimer’s disease \citep{risacher2009baseline,lee2015bflcrm,kong2018flcrm}. Alzheimer's disease (AD) is a progressive neurodegenerative disease affecting more than 
57 million people worldwide \citep{gaugler20192019,who2025dementia} and is the most common form of dementia. In the United States, over 7.2 million Americans aged 65 and older are living with AD as of 2025 \citep{alzheimer2025}, and cases are projected to more than double by 2050 \citep{hebert2013alzheimer} in the United States. The absence of any permanent treatment to cure AD makes early detection, identification of biomarkers, and predicting time-to-conversion to AD of paramount importance. Subtle distributional differences in ROI-level intensities may carry important prognostic information in AD studies, which is lost under mean-based summaries. Prior work has examined structural MRI measures, including regional
volume \citep{liu2010analysis}, diffusion-based metrics
\citep{acosta2012diffusion}, and cortical thickness
\citep{phan2024increased} to characterize neurodegeneration and model
conversion from mild cognitive impairment to AD.
These approaches use summaries of voxel-level measurements within
anatomically defined ROIs and thus discard the within-region
heterogeneity.
More recent machine learning approaches incorporate high-dimensional
imaging features \citep{wang2021deep}, but these often treat voxels
or regions independently, and may lack interpretability and stability
in time-to-event analysis.
Our proposed approach overcomes these limitations by retaining
distributional information at the ROI level while incorporating
structural dependence across regions within a survival framework.



From a modeling perspective, these considerations naturally fall within the framework of distributional data analysis (DDA), an evolving area of Statistics \citep{petersen2022modeling} useful for capturing subject-specific distributions through histograms, densities, quantile functions, or other distributional representations and incorporating them as observations within different statistical modeling techniques. 
Several distributional regression models have been proposed depending on whether the outcome or the predictor is distribution-valued. For scalar outcomes with distributional predictors, approaches include scalar-on-quantile function regression \citep{ghosal2021scalar,ghosal2023distributional,zhang2024scalar}, kernel-based nonparametric methods \citep{matabuena2020glucodensities,matabuena2023distributional}, and density-transformation based frameworks \citep{petersen2016functional, talska2021compositional}. 
 Regression approaches have also been proposed 
 for modeling distribution-valued objects as outcome 
 using Fr\'echet regression and shape-constrained approaches, which can accommodate Euclidean \citep{petersen2019frechet,yang2020quantile,yang2020random,bhattacharjee2023single,matabuena2025predicting,matabuena2024multilevel,bhattacharjee2025geodesic}, distributional \citep{chen2021wasserstein,ghodrati2022distribution,pegoraro2022projected}, and a combination of scalar, distributional \citep{ghosal2025distributional} predictors. However, none of these frameworks is designed for censored
time-to-event outcomes or for settings with high-dimensional
graph-structured distributional predictors.

\begin{figure}[ht]
    \centering
    \includegraphics[width=\textwidth]{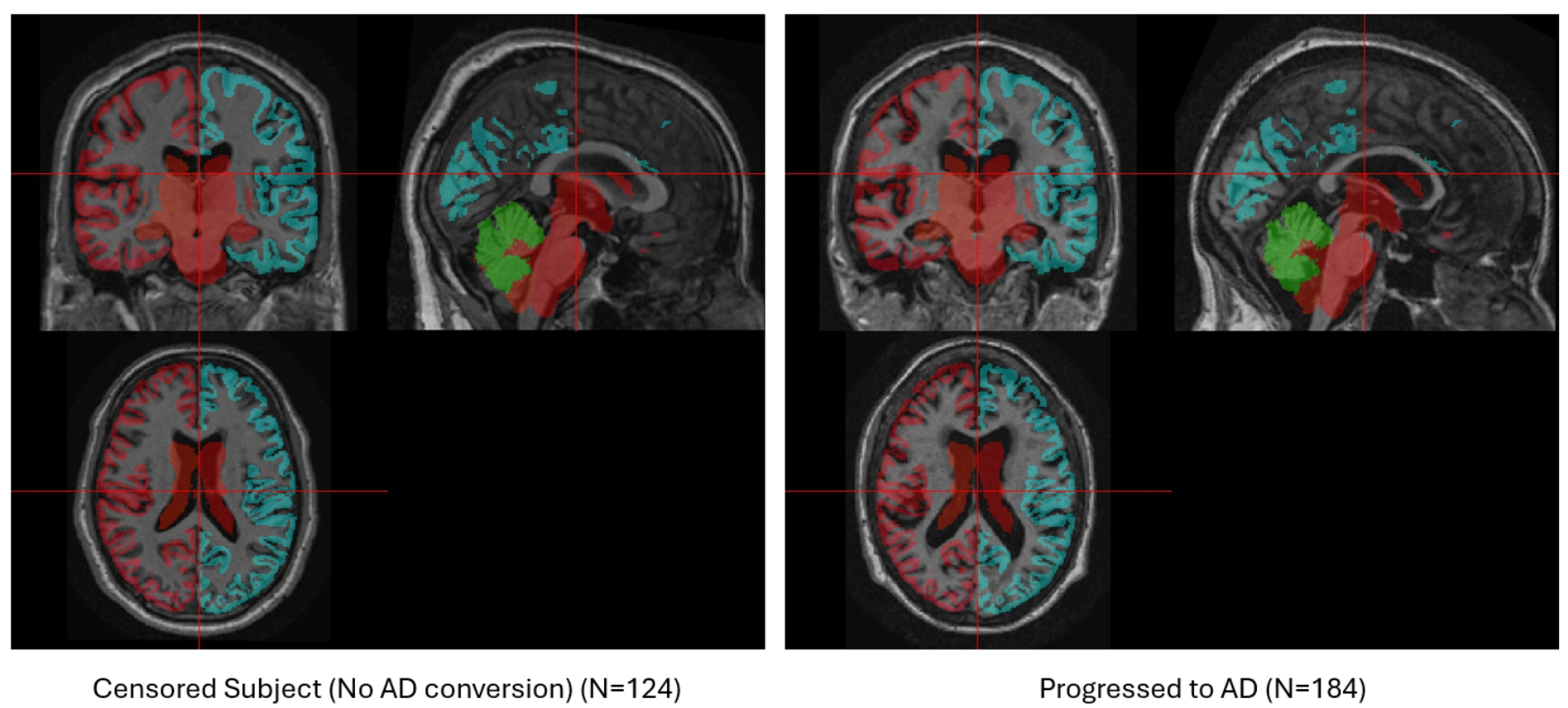}
    \caption{Orthogonal views of baseline T1-weighted MRI with DKT atlas overlays for representative ADNI subjects. The left panel corresponds to a censored subject (no AD conversion, N = 124), and the right panel to a subject who progressed to AD (N = 184). Colored regions denote atlas-defined ROIs used for distributional modeling.}
    \label{fig:adniImage}
\end{figure}

In this article, our motivating application is based on a neuroimaging dataset from the Alzheimer’s Disease Neuroimaging Initiative (ADNI). 
Figure \ref{fig:adniImage} illustrates representative baseline T1-weighted MRI scans with Desikan-Killiany-Tourville (DKT) atlas overlayed for two representative subjects (one progressed to AD and one censored), highlighting the anatomical parcellation of the brain into regions that will serve as the basis for ROI-level distributional representations in our analysis.


\subsection{Related Work}
Despite substantial advances in distributional data analysis, several computational and statistical challenges remain when analyzing high-dimensional neuroimaging data for disease prognosis, particularly when multiple ROI-level distributional measurements are involved. First, most existing distributional regression frameworks do not directly accommodate censored time-to-event outcomes, which are central in studies of conversion to Alzheimer’s disease, such as our motivating study. Second, the direct incorporation of distributional representation into a functional Cox model \citep{gellar2015cox,lee2015bflcrm,kong2018flcrm,cui2021additive,yang2021weighted} does not address the high-dimensional setting in which numerous ROI-level distributions must be considered simultaneously, necessitating principled regularization and region selection. Third, these methods generally treat distribution-valued predictors independently and do not exploit known spatial or graphical relationships among ROIs, which may be essential for stable estimation and biological interpretability in neuroimaging studies. Together, these limitations highlight the need for a unified modeling framework that integrates distribution-valued predictors into survival analysis while simultaneously achieving selection and fusion of ROIs based on known dependency structures.

A substantial body of work has established sparsity-inducing regularization as a central tool for high-dimensional regression and survival analysis. The LASSO and related $\ell_1$-penalized estimators were extended to the Cox proportional hazards model to enable variable selection under censoring \citep{tibshirani1997lasso, huang2013oracle}, providing stable estimation in high-dimensional settings. Beyond simple sparsity, structured penalties have been developed to exploit known relationships among predictors. The fused LASSO \citep{tibshirani2005sparsity} introduced joint penalties on coefficient magnitudes and pairwise differences to encourage local similarity, and the generalized LASSO framework \citep{roth2004generalized} extended this idea to arbitrary graph structures. Network-constrained regularization and graph-guided penalties incorporate prior connectivity information, often via graph Laplacians or fusion penalties \citep{friedman2008sparse, mazumder2012graphical,cuingnet2012spatial,beer2019incorporating,ranciati2021fused}, and have been widely applied in classical linear and generalized linear models, particularly in neuroimaging and genomics analysis. These ideas were subsequently adapted to survival analysis, including network-regularized Cox models \citep{verissimo2016degreecox, sun2014network} and fused-LASSO–type Cox formulations \citep{zhou2012modeling, sennhenn2016structured} that borrow strength across spatially adjacent features. Parallel to this literature, functional Cox models were developed to accommodate functional covariates \citep{gellar2015cox,qu2016optimal,cui2021additive} varying over a continuous domain, where coefficient functions are estimated semi-parametrically using penalized splines. More recent work has focused on variable selection in a high-dimensional functional Cox model using group sparsity penalties to select the relevant functional predictors \citep{yue2025variable,chen2025high}. However, these two strands of literature remain largely separate. Graph-regularized Cox models primarily address scalar covariates with network structure, while the existing estimation and variable selection methods in the functional Cox model typically treat functional predictors independently and focus on smoothness within each function rather than structural relationships across multiple functional units. In neuroimaging applications, where each ROI is represented by a distributional object, and ROIs are anatomically connected, neither framework alone is sufficient. There remains a  clear methodological gap for a survival analysis framework that simultaneously performs interpretable region selection, incorporates graph-based fusion across related ROIs, and accommodates distribution-valued predictors. 

\subsection{Our Contributions}
\begin{figure}[h!]
    \centering
    \includegraphics[width=\linewidth]{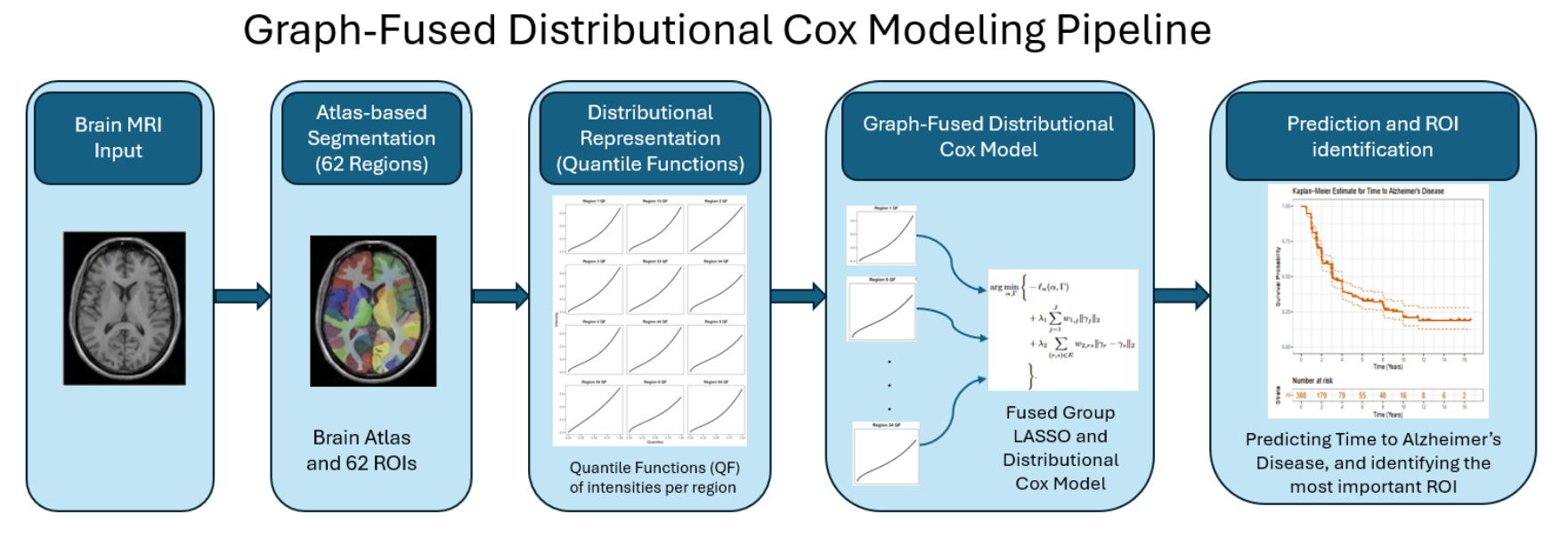}
    \caption{The schematic pipeline for the proposed graph-fused distributional Cox model (GFDCM).}
    \label{fig:placeholder}
\end{figure}

We address these challenges by proposing the Graph-Fused Distributional Cox Model (GFDCM), a unified framework that integrates distribution-valued predictors, structured regularization and fusion, and a time-to-event outcome, a setting which cannot be accommodated by existing methods. Our contributions are threefold: i) GFDCM incorporates node-level quantile functions directly as predictors within the Cox proportional hazards model, preserving within-region distributional heterogeneity rather than reducing voxel intensities to scalar summaries, 
 ii) GFDCM introduces a structured regularization scheme that combines sparsity and graph-based fusion to achieve simultaneous variable selection and information sharing across graphically connected nodes. The sparsity component identifies a subset of prognostically relevant
nodes, while the graph-fusion penalty encourages similar coefficient functions among neighboring nodes, improving estimation stability and interpretability in high-dimensional settings, iii) We develop an efficient optimization algorithm based on the alternating direction method of multipliers (ADMM) to solve the resulting penalized partial likelihood problem, and we implement a data-driven tuning parameter selection strategy using information criteria to balance goodness of fit and model complexity.  A schematic representation of the proposed framework is given in Figure \ref{fig:placeholder}.


The remainder of the paper is organized as follows. In Section \ref{sec:2}, we present the proposed methodology. We begin in Section \ref{sec:21} by introducing the overall modeling framework and notation. Section \ref{sec:22} develops the Distributional Cox Model. Section \ref{sec:23} describes the estimation strategy, followed by Section \ref{sec:24}, which introduces the graph-guided fusion penalty. Section \ref{sec:25} details the optimization procedure based on the alternating direction method of multipliers (ADMM), and Section \ref{sec:26} discusses the data-driven selection of tuning parameters. Section \ref{sec:3} presents simulation studies designed to evaluate estimation accuracy, variable selection performance, and robustness of the proposed method. Finally, Section \ref{sec:4} applies the proposed method to ADNI data to identify prognostic brain regions and their distributional representations associated with Alzheimer’s disease conversion. We conclude in Section \ref{disc} with a discussion on the contribution and possible extensions.

\section{Methodology}
\label{sec:2}
\subsection{Modelling Framework}
\label{sec:21}

We denote by $Q_{ij}(p)$ ($i=1,\ldots,n$ and $j=1,\ldots,J$) the latent subject and node-specific quantile function \citep{ghosal2023distributional,ghosal2025distributional} representing the distribution of measurements for subject $i$ at node
$j$ of the graph $G=(V,E)$, where $V=\{1,\ldots,J\}$ indexes nodes
and $(r,s)\in E$ if nodes $r$ and $s$ are connected. 
For example, in neuroimaging data, the nodes represent the regions of interest (ROIs), and they are based on a standard anatomical atlas, which divides the brain into $J$ distinct ROIs. For each subject $i$ and node $j$, we observe a collection of
measurements $W_{ijk}$ $(k=1,\ldots,n_{ij})$, assumed to come from
a latent distribution with c.d.f.\ $F_{ij}$ and corresponding
quantile function $Q_{ij}(p)=\inf\{x:F_{ij}(x)\geq p\}$.
In practice, the true latent quantile function $Q_{ij}(p)$ is not
directly observed, it is estimated from
$\{W_{ijk}\}_{k=1}^{n_{ij}}$ via the empirical c.d.f.\
$\hat{F}_{ij}(x)=n_{ij}^{-1}\sum_{k=1}^{n_{ij}}\mathbf{1}\{W_{ijk}
\leq x\}$ and linear interpolation of order statistics \citep{yang2020random,ghosal2025distributional}.
The empirical quantile function
$\widetilde{Q}_{ij}(p)$ for
$p\in[\frac{1}{n_{ij}+1},\frac{n_{ij}}{n_{ij}+1}]$ is given by
\citep{hyndman1996sample}:
\begin{equation}
  \widetilde{Q}_{ij}(p)
  = (1-w)\,W_{(ij\lfloor(n_{ij}+1)p\rfloor)}
    + w\,W_{(ij\lfloor(n_{ij}+1)p\rfloor+1)},
  \label{eq:quantile_est}
\end{equation}
where $w$ satisfies $(n_{ij}+1)p=\lfloor(n_{ij}+1)p\rfloor+w$.
In practice, we evaluate this on an equi-spaced grid of quantiles $\mathcal{P} = \{p_1, \dots, p_m\} \subset [0,1]$. 
This distributional representation completely characterizes the
within-node variation, naturally accommodates differing numbers of measurements across nodes and subjects, and retains tail information that is lost under mean or median-based scalar summaries.

Let $T_i$ be the survival time of interest for subject $i$, and $C_i$ the corresponding censoring time for 
$i=1,\ldots,n$. The observed event time in the presence of right censoring is given by $Y_i= \min(T_i,C_i)$, and $\Delta_i= I(T_i\leq C_i)$ is the event or censoring indicator. Let $\*Z_i=(Z_{i1},\ldots,Z_{iq})^T$ denote  scalar
baseline covariates. The data for the $i$th subject are
$\mathbf{D}_i=(Y_i,\Delta_i,\*Z_i,Q_{i1}(\cdot),\ldots,
Q_{iJ}(\cdot))$, $i=1,\ldots,n$, assumed to be independent and
identically distributed.
We assume that $T_i\perp C_i\mid(\*Z_i,Q_{i1}(\cdot),\ldots,
Q_{iJ}(\cdot))$ (independent censoring).
The graph $G=(V,E)$ encodes known structural, spatial, or functional
relationships among the $J$ nodes; in practice it may represent
anatomical adjacency, pathway membership, or any other domain-specific
connectivity. Since the true latent $Q_{ij}(p)$ is not
directly observed, we will plug in the estimated $\widetilde{Q}_{ij}(p)$ in the models posited throughout the article \citep{matabuena2023distributional,ghosal2023distributional,ghosal2025distributional} during estimation.

\subsection{Distributional Cox Model}
\label{sec:22}
We model the association between the node-level distributions and a right-censored time-to-event outcome using a distributional Cox model, which extends the functional Cox model of \citet{gellar2015cox,kong2018flcrm,cui2021additive} to the setting of high-dimensional, graph-structured distribution-valued predictors. The hazard function for subject $i$ at time $t$ is modeled as:
\begin{equation}
\label{eq:cox_model1}
\lambda_i(t \mid \*Z_i,\{Q_{ij}(\cdot)\}_{j=1}^J) = \lambda_0(t) \exp\left( \*Z_i^T\bm\alpha+ \sum_{j=1}^J \int_0^1 Q_{ij}(p) \beta_j(p) \, dp \right),
\end{equation}
where $\lambda_0(t)$ is an unspecified baseline hazard function and $\bm\alpha$ quantifies the association with the scalar covariates.  The node-specific
distributional coefficient function $\beta_j(p)$ captures how different quantile levels of the distribution at node $j$ relate to the log-hazard. The function $\beta_j(p)$ assigns differential importance to
different quantile levels, lower tails, medians, or upper
tails, in a fully data-driven manner.
When $\beta_j(p)=\beta_j$ is constant in $p$, model
\eqref{eq:cox_model1} reduces to a standard proportional hazards model with
mean-valued predictors $\bar Q_{ij}=\int_0^1 Q_{ij}(p)\,dp$, making
it a strict generalization. The distributional coefficient functions $\beta_j(\cdot)$ ($j=1,\ldots,J$) are assumed to be smooth functions with finite second derivative.


\subsection{Estimation }
\label{sec:23}
The unknown parameters of interest in the above model, which need to be estimated, are given by $\bm\Theta=(\bm\alpha,\beta_1(\cdot),\ldots,\beta_J(\cdot),h_0(t))$. We model the coefficient functions $\beta_j(\cdot)$ using basis function expansions as $\beta_j(p) = \sum_{k=1}^d \theta_{jk} B_k(p) = \mathbf{B}(p)^\top \boldsymbol{\theta}_j,$ 
where $\{B_k(p)\}_{k=1}^d$ is a set of known basis functions on $[0,1]$ and $\boldsymbol{\theta}_j = (\theta_{j1}, \dots, \theta_{jd})^\top$ is the vector of unknown basis coefficients for node $j$. In this article, we use cubic B-spline basis functions; however, other basis functions can be used as well. Plugging in this basis expansion of the coefficient functions in (2), we have the linear predictor given by,
\begin{equation}
    \eta_i =  \*Z_i^T\bm\alpha + \sum_{j=1}^J \boldsymbol{\theta}_j^T \left( \int_0^1 Q_{ij}(p) \mathbf{B}(p) \, dp \right) = \*Z_i^T\bm\alpha + \sum_{j=1}^J \mathbf{W}_{ij}^T {\bm\theta}_j,
\end{equation}
where $\mathbf{W}_{ij}$ is the $d$-dimensional vector of projected covariates $ \int_0^1 Q_{ij}(p) \mathbf{B}(p) \, dp$ for node $j$. 

To ensure smooth and sparse interpretable coefficient functions, we impose a simultaneous sparsity and smoothness inducing penalty on the coefficient functions \citep{ghosal2020variable,yue2025variable} as:
\begin{equation}\label{eq:6}
\begin{split}
P_{\lambda,\psi}\{\beta_j (\cdot)\} &= \lambda \left\{ \int \beta_j(\cdot)^2 \, dp + \psi \int \beta_j''(\cdot)^2 \, dp \right\}^{1/2} \\
&= \lambda \left( \boldsymbol{\theta}_j^T \mathbb{R} \boldsymbol{\theta}_j + \psi \boldsymbol{\theta}_j^T \mathbb{Q} \boldsymbol{\theta}_j \right)^{1/2} = \lambda \left( \boldsymbol{\theta}_j^T \mathbb{K}_{\psi} \boldsymbol{\theta}_j \right)^{1/2}.
\end{split}
\end{equation}
Here \(\mathbb{R} = \int \mathbf{B}(p) \mathbf{B}(p)^T dp\), and \(\mathbb{Q} = \int \mathbf{B}''(p) \mathbf{B}''(p)^T \, dp\) and \(\mathbb{K}_{\psi} = \mathbb{R} + \psi \mathbb{Q}\). The smoothing parameter \(\psi \geq 0\) controls the roughness of the coefficient function. We use a reparameterization of the linear predictor using the Cholesky decomposition \(\mathbb{K}_{\psi} = \mathbb{L}_{\psi} \mathbb{L}_{\psi}^T\) \citep{yue2025variable,ghosal2020variable} and define \(\boldsymbol{\gamma}_j = \mathbb{L}_{\psi}^T \boldsymbol{\theta}_j\), $\mathbf{U}_{ij}=\mathbb{L}_{\psi}^{-1}\mathbf{W}_{ij}$.  The linear predictor can be rewritten as 
\begin{equation}
     \eta_i =  \*Z_i^T\bm\alpha + \sum_{j=1}^J \mathbf{U}_{ij}^T \boldsymbol{\gamma}_j,
\end{equation}
and the penalty (4) reduces to $P_{\lambda,\psi}\{\beta_j (\cdot)\}=\lambda(\boldsymbol{\gamma}_j^T \boldsymbol{\gamma}_j )^{1/2}=\lambda \|\boldsymbol{\gamma}_j\|_2$, which resembles a group LASSO penalty \citep{yuan2006model} on the basis coefficients, and can impose group-sparsity to achieve node selection. The penalty imposes a group-level $\ell_2$ norm
on the coefficient vector, so that $\bm\gamma_j=\mathbf{0}$
removes node $j$ entirely from the model.

\subsection{Graph-Fused Group LASSO Penalty}
\label{sec:24}
In many applications, connected nodes in $G$ are expected to have similar or identical effects on the outcome.
For example, adjacent brain regions tend to exhibit correlated
structural changes, and nearby genes share regulatory pathways.
We exploit this prior structural knowledge, encoded in $G=(V,E)$, to guide variable selection through a penalization framework that
simultaneously performs selection of nodes (e.g., identifying ROIs associated with AD) and spatial fusion (identifying and fusing clusters of graph-connected nodes with identical distributional effects). We define the Graph-Fused Group LASSO (GFGL) penalty as:

\begin{equation}
\label{eq:penalty}
P(\boldsymbol{\Gamma}) = \lambda_1 \sum_{j=1}^J w_{1,j} \|\boldsymbol{\gamma}_j\|_2 + \lambda_2 \sum_{(r,s) \in E} w_{2,rs} \|\boldsymbol{\gamma}_r - \boldsymbol{\gamma}_s\|_2,
\end{equation}
where $\boldsymbol{\Gamma} = (\boldsymbol{\gamma}_1^\top, \dots, \boldsymbol{\gamma}_J^\top)^\top$ consists of all the basis coefficients. 
The first term is a group LASSO penalty which induces sparsity at the node level; if $\|\boldsymbol{\gamma}_j\|_2 = 0$, the entire distributional effect of node $j$ is removed from the model. The second term works as a graph-fused group LASSO penalty. It encourages the coefficient vectors of adjacent nodes $r$ and $s$ to be identical ($\boldsymbol{\gamma}_r = \boldsymbol{\gamma}_s$), effectively fusing them if their effects on the log-hazard are similar. We denote by $w_{1,j}$ and $w_{2,rs}$ the adaptive weights, which might be chosen based on an initial fit and prior information, e.g., the distance between the connected nodes. The partial likelihood function for the distributional Cox model (2) and reparametrized linear predictor (5) is given by:
 \begin{equation}
L(\boldsymbol{\alpha}, \boldsymbol{\Gamma}) = \prod_{d \in D} \frac{\exp(\eta_d)}{\sum_{i \in R_d} \exp(\eta_i)},
\end{equation}
where \(D\) is the set of failure indices, \(R_d\) is the set of individuals at risk at time \(t_d\). Let us denote the log partial likelihood by: $\ell_n(\boldsymbol{\alpha}, \boldsymbol{\Gamma}) = n^{-1} \log L(\boldsymbol{\alpha}, \boldsymbol{\Gamma})$.
The unknown parameters are estimated by minimizing the negative partial log-likelihood regularized by the GFGL penalty:
\begin{eqnarray}
   (\hat{\boldsymbol{\alpha}}, \hat{\boldsymbol{\Gamma})} &=\underset{\bm\alpha, \bm\Gamma}{\arg \min} \left\{ -\ell_n(\boldsymbol{\alpha}, \boldsymbol{\Gamma}) + P(\boldsymbol{\Gamma}) \right\}\notag\\
   &=\underset{\bm\alpha, \bm\Gamma}{\arg \min} \left\{ -\ell_n(\boldsymbol{\alpha}, \boldsymbol{\Gamma}) + \lambda_1 \sum_{j=1}^J w_{1,j} \|\boldsymbol{\gamma}_j\|_2 + \lambda_2 \sum_{(r,s) \in E} w_{2,rs} \|\boldsymbol{\gamma}_r - \boldsymbol{\gamma}_s\|_2 \right\}.
\end{eqnarray}
The optimization of the above GFGL penalized Cox model (8) is non-standard due to the non-differentiability and non-separability of the graph-fused group LASSO penalty (second-term). Although variants of the above optimization have been proposed with the least-squares loss function \citep{beer2019incorporating}, this work is among the first to integrate the group-LASSO and graph-fused group-LASSO
penalties for graph-guided fusion and variable selection in the Cox model with distribution-valued predictors at graph nodes. We employ the alternating direction method of multipliers (ADMM) algorithm for the above optimization problem \citep{boyd2011distributed}.

\subsection{Optimization via ADMM}
\label{sec:25}
We rewrite the penalized objective function in (8) as
\begin{eqnarray}
   (\hat{\boldsymbol{\alpha}}, \hat{\boldsymbol{\Gamma})} 
   &=\underset{\bm\alpha, \bm\Gamma}{\arg \min} \left\{ -\ell_n(\boldsymbol{\alpha}, \boldsymbol{\Gamma}) + \lambda_1 \sum_{j=1}^J w_{1,j} \|\boldsymbol{\gamma}_j\|_2 + \lambda_2 \sum_{(r,s) \in E} w_{2,rs} \|\boldsymbol{\gamma}_r - \boldsymbol{\gamma}_s\|_2 \right\} \notag\\
    &= \underset{\bm\alpha, \bm\Gamma}{\arg \min} \left\{ -\ell_n(\boldsymbol{\alpha}, \boldsymbol{\Gamma}) + \sum_{l=1}^{N} \lambda_lw_l\|\^K_l\boldsymbol{\Gamma}\|_2\right\}.
\end{eqnarray}

Here, $N = J + |E|$ represents the total number of groups, where $J$ is the number of nodes and $|E|$ is the number of edges in the connectivity graph (note that $|E| = J(J-1)/2$ in the case of a fully connected graph), and we have
\begin{equation}
    (\lambda_l, \^K_l) = 
    \begin{cases} 
        (\lambda_1, \^G_l) & \text{for } l = 1, \dots, J \quad (\text{selection}), \\
        (\lambda_2, \^A_{l}) & \text{for } l = J+1, \dots, N \quad (\text{graph fusion}),
    \end{cases}
\end{equation}
Here $\^G_l$ (for  $l=1,\dots,J$) is the matrix for group-LASSO, a sparse $d \times Jd$ block matrix that extracts the coefficients for the $l$-th node. It consists of a $d \times d$ identity matrix $\^I_d$ at the $l$-th block position and zeros elsewhere,
\[
    \^{G}_l = \left[ \mathbf{0}, \dots, \underbrace{\^I_d}_{l\text{-th block}}, \dots, \mathbf{0} \right].
    \]

$\^A_l$ are pairwise difference matrices $\^A_{r,s}$ introducing fusion between node $r$ and node $s$ coefficients. For each edge $e=(r,s) \in E$, $\^A_{r,s}$ is a $d \times Jd$ matrix that computes the difference between the coefficient vectors of nodes $r$ and $s$. It places $\^I_d$ at the $r$-th block and $-\^I_d$ at the $s$-th block:

 \[
    \^{A}_{rs} = \left[ \mathbf{0}, \dots, \underbrace{\^{I}_d}_{r\text{-th}}, \dots, \underbrace{-\^{I}_d}_{s\text{-th}}, \dots, \mathbf{0} \right].
    \]

Finally, $w_l$ are known weights for group-LASSO and graph-fused group LASSO, respectively.  Now, to solve the above optimization problem, we use an ADMM algorithm. For ADMM, we introduce the auxiliary variables $\boldsymbol{\theta}_l = \^K_l \boldsymbol{\Gamma}$. The optimization problem becomes,
\begin{equation}
    (\hat{\boldsymbol{\alpha}}, \hat{\boldsymbol{\Gamma})} 
   =\underset{\bm\alpha, \bm\Gamma}{\arg \min} \left\{ -\ell_n(\boldsymbol{\alpha}, \boldsymbol{\Gamma}) +   \sum_{l=1}^N \lambda_l w_l \| \boldsymbol{\theta}_l \|_2\right\}
 , \quad \text{subject to } \boldsymbol{\theta}_l - \^K_l \boldsymbol{\Gamma} = 0 \text{ for } l \in \{1, 2, \ldots, N\}.
\end{equation}
 Let us denote $\^K = (\^K_1 | \ldots | \^K_N)$, $\boldsymbol{\theta} = (\boldsymbol{\theta}_1, \ldots, \boldsymbol{\theta}_N)^T$, and $\boldsymbol{\mu} = (\boldsymbol{\mu}_1, \ldots, \boldsymbol{\mu}_N)^T$. The augmented Lagrangian for the above objective function is given by,
\begin{equation}
    \mathcal{L}_\rho(\bm\alpha,\boldsymbol{\Gamma}, \boldsymbol{\theta}, \boldsymbol{\mu}) = -\ell_n(\boldsymbol{\alpha}, \boldsymbol{\Gamma})+ \sum_{l=1}^N \lambda_l w_l \| \boldsymbol{\theta}_l \|_2 + \sum_{l=1}^N \boldsymbol{\mu}_l^T (\boldsymbol{\theta}_l - \^K_l \boldsymbol{\Gamma}) + \frac{\rho}{2} \| \boldsymbol{\theta}_l - \^K_l \boldsymbol{\Gamma} \|_2^2.
\end{equation}


Here $\rho > 0$ is the step-size parameter and $\boldsymbol{\mu}_l$ are the dual variables for ADMM. After initialization of $\bm\alpha,\boldsymbol{\Gamma}$, $\boldsymbol{\theta}$, and $\boldsymbol{\mu}$, the ADMM algorithm iterates through the following three update steps until convergence:

\paragraph{Primal update ($\bm\alpha,\bm\Gamma$).}
The primal update is given by 
\begin{equation}
    (\alpha^{t+1},\boldsymbol{\Gamma}^{t+1}) = \arg \min_{\boldsymbol{\alpha,\Gamma} } \mathcal{L}_\rho(\bm\alpha,\boldsymbol{\Gamma}, \boldsymbol{\theta}^t, \boldsymbol{\mu}^t). \\
\end{equation}
Note that this is a convex optimization problem involving the negative Cox partial likelihood and a quadratic (and linear) penalty term, and hence has a global minimum. In our implementation, the primal update is solved numerically by a quasi-Newton (BFGS) method.

\paragraph{Auxiliary update ($\bm\theta$).}
The auxiliary variable update is given by
\begin{equation}
    \boldsymbol{\theta}^{t+1} = \arg \min_{\boldsymbol{\theta}} \mathcal{L}_\rho(\bm\alpha^{t+1},\boldsymbol{\Gamma}^{t+1}, \boldsymbol{\theta}, \boldsymbol{\mu}^t). \\
\end{equation}
Note that for the $\boldsymbol{\theta}^{t+1}$ updates, the corresponding $\mathcal{L}_\rho$ subgradient will equal zero at the optimal solution. Hence, based on the proximal operator for group LASSO (which is a group soft-thresholding operator), the update for  $\boldsymbol{\theta}^{t+1}$ is given by (for $l=1,\ldots,N$)
:

\begin{equation}
    \boldsymbol{\theta}_l^{(t+1)} = \mathcal{S}\left( \^{K}_l \boldsymbol{\Gamma}^{(t+1)} - \rho^{-1}\boldsymbol{\mu}_l^{(t)}, \, \frac{\lambda_l w_l}{\rho} \right),
\end{equation}
where the group soft-thresholding operator is defined as $\mathcal{S}(\mathbf{v}, \tau) = \left(1 - \frac{\tau}{\|\mathbf{v}\|_2}\right)_+ \mathbf{v}$, and $(x)_+ = \max(0, x)$.

\paragraph{Dual update ($\bm\mu$).}
 \begin{equation}
     \boldsymbol{\mu}_l^{t+1} = \boldsymbol{\mu}_l^t + \rho(\boldsymbol{\theta}_l^{t+1} - \^K_l \boldsymbol{\Gamma}^{t+1}), \hspace{2 mm} \textit{for} \hspace{2 mm} l=1,\ldots N. 
 \end{equation}
\paragraph{Stopping criteria.}
 We use the stopping criteria recommended in \cite{boyd2011distributed}, and stop the algorithm 
when the primal and dual residuals are smaller than a linear combination of
preselected levels of  $\epsilon^{\text{abs}}$ and 
relative tolerance $\epsilon^{\text{rel}} $. The primal ($\*{r}$) and dual ($\mathbf{s}$) residuals at the current stage are calculated as:
\begin{eqnarray}
    \*{r}^{(t)} = \left[ (\^{K}_1 \boldsymbol{\Gamma}^{(t)} - \boldsymbol{\theta}_1^{(t)})^\top, \dots, (\^{K}_N \boldsymbol{\Gamma}^{(t)} - \boldsymbol{\theta}_N^{(t)})^\top \right]^\top.\\
    \mathbf{s}^{(t)} = \rho \sum_{l=1}^N \mathbf{K}_l^\top \left( \boldsymbol{\theta}_l^{(t)} - \boldsymbol{\theta}_l^{(t-1)} \right).
\end{eqnarray}
    Convergence is achieved when $\|\mathbf{r}^{(t)}\|_2 \leq \epsilon^{\text{pri}}_{(t)}$ and $\|\mathbf{s}^{(t)}\|_2 \leq \epsilon^{\text{dual}}_{(t)}$ , where the tolerance thresholds are defined as:

\begin{eqnarray}
      \epsilon^{\text{pri}}_{(t)} &= \epsilon^{\text{abs}} \sqrt{Nd} + \epsilon^{\text{rel}} \max \left( \sqrt{\sum_{l=1}^N \|\mathbf{K}_l \boldsymbol{\Gamma}^{(t)}\|_2^2}, \, \sqrt{\sum_{l=1}^N \|\boldsymbol{\theta}_l^{(t)}\|_2^2} \right), \\
    \epsilon^{\text{dual}}_{(t)} &= \epsilon^{\text{abs}} \sqrt{Jd} + \epsilon^{\text{rel}} \left\| \sum_{l=1}^N \mathbf{K}_l^\top \boldsymbol{\mu}_l^{(t)} \right\|_2.
\end{eqnarray}
The absolute and relative tolerance parameters $\epsilon^{\text{abs}}$ and $\epsilon^{\text{rel}}$ are set to $10^{-4}$ in our implementation throughout this article. 

\subsection{Tuning Parameter Selection}
\label{sec:26}

The proposed GFDCM method relies on four tuning parameters: the smoothness parameter $\psi$, sparsity penalty parameter $\lambda_1$, the fusion penalty parameter $\lambda_2$, and the basis dimension $d$. Directly doing a four-dimensional grid search over the tuning-parameter space is computationally costly. Hence, we adopt a hybrid strategy in this article to select the tuning parameters, which strikes a balance between computational efficiency and model performance. First, for a fixed $\psi$ and number of basis $d$, we choose the sparsity penalty parameter  $\lambda_1$ using a group-MCP penalized version of the objective function (9) (without any fusion penalty) and the EBIC criterion \citep{chen2008extended}. This is denoted as $\lambda_1(\psi,d)$. We use the computationally efficient implementation proposed in \cite{yue2025variable} to achieve this. Now, for a given $\psi,d,\lambda_1(\psi,d)$, we fix a $\lambda_2$, and  fit the GFDCM model via the proposed ADMM algorithm and compute the following Bayesian Information Criterion (BIC), adapted for the effective degrees of freedom in fused models:
\begin{equation}
    \text{BIC}(\psi,d,\lambda_1(\psi,d), \lambda_2) = -2n\ell_n(\hat{\boldsymbol{\alpha}}, \hat{\boldsymbol{\Gamma}}) + \nu d\log(n),
\end{equation}
where $\nu$ represents the number of unique functional groups identified (i.e., the count of distinct non-zero coefficient vectors after fusion). 
The optimal values of the tuning parameters $(\psi,d,\lambda_1(\psi,d), \lambda_2)$ are empirically chosen to minimize this BIC criterion over a grid. For the step-size $\rho$, we set $\rho=1$ in this article, which has shown good empirical performance. Alternatively, an adaptive step size \citep{beer2019incorporating} can be set to accelerate the convergence of the algorithm.

\section{Simulation Study}
\label{sec:3}
In this Section, we investigate the estimation, selection, and fusion performance of the proposed GFDCM approach through numerical simulations. We consider two scenarios that share the same data-generating model. In Scenario A, the graph supplied to GFDCM coincides with the true dependency structure among the nodes (i.e. the graph is known). We explore an additional Scenario B, in which the supplied graph is misspecified, which allows us to assess the robustness of GFDCM to errors in the assumed graph.


\subsection{Simulation Design}
\paragraph{Scenario A (correctly specified graph).}

 We generated survival times from the following distributional Cox model:
\begin{equation}
\log\lambda_i(t \mid \*Z_i,\{Q_{ij}(\cdot)\}_{j=1}^J)= \log \lambda_0(t) + \mathbf{Z}_i^\top \boldsymbol{\alpha} + \sum_{j=1}^{J} \int_{0}^{1} Q_{ij}(p) \beta_j(p) dp, 
\end{equation}
for $i=1,\ldots,n$.  Here $\lambda_0(t)$ denotes the baseline hazard function, $\mathbf{Z}_i = (Z_{i1}, Z_{i2})^\top$ are scalar covariates, and $\{Q_{ij}(p)\}_{j=1}^J$ are the node-level distributional predictors for $J=10$ nodes (e.g., ROIs in our application). The scalar covariates $Z_{i1}$ and $Z_{i2}$, were generated independently from a Uniform distribution $U(-1, 1)$. To mimic the non-decreasing property of quantile functions, the distributional predictors $Q_{ij}(p)$ were generated using a monotone B-spline basis expansion $Q_{ij}(p)=\sum_{k=1}^{52}\zeta_{ijk} B_k(p)$. Specifically, we used cubic B-splines where the basis coefficients $\zeta_{ijk}$ were simulated from a cumulative sum of Gamma random variables $\zeta_{ijk}=\sum_{l=1}^{k}G_{ijl}$, where $G_{ijl}\sim Gamma (0.2,0.2)$. This ensures $\zeta_{ij,k}< \zeta_{ij,k+1}$ and leads to non-decreasing $Q_{ij}(p)$  with respect to $p$. We used $m=51$ equally-spaced grid points $p_l \in [0, 1],l=1,\ldots,51$, on which $Q_{ij}(p)$ were observed. For the known dependency structure among the nodes (regions), we assume nodes 1 and 2 are connected, and they share an identical non-zero effect
$\beta_1(p) = \beta_2(p) = 5 \cos(2\pi p)$. Similarly, we assume nodes 3, 4, and 5 are connected, and these nodes share a different non-zero effect:
   $\beta_3(p) = \beta_4(p) = \beta_5(p) = 3 \sin(\pi p)$.
The remaining five nodes have no effect on survival, i.e., $\beta_j(p) = 0$ for $j = 6, \dots, 10$, and form two connected components among themselves: nodes 6 and 7 are connected, and nodes 8, 9, and 10 are mutually connected. The resulting graph has edge set
\[
\mathcal{E} = \{(1,2), (3,4), (3,5), (4,5), (6,7), (8,9), (8,10), (9,10)\},
\]
i.e., $|\mathcal{E}| = 8$, and contains no edges between the active nodes ($1$-$5$) and the null nodes ($6$-$10$).   
This graph structure implies that a successful method should identify two distinct active groups and correctly zero out the coefficients for the remaining nodes. We set the scalar coefficients $\boldsymbol{\alpha} = (0.5, 1)^T$. The true survival times $T_i$ were generated from an exponential distribution with subject-specific hazard rates $\lambda_i = \exp(\eta_i)$, where $\eta_i$ is the linear predictor. Censoring times $C_i$ were generated independently from an exponential distribution with mean $\mu_c=1$, leading to a $20\%$ censoring rate on average.
We consider sample sizes \(n \in \{300, 500, 1000\}\) and generate $M =100$ Monte Carlo (M.C) replications from each scenario to assess the performance of the proposed GFDCM.
\paragraph{Scenario B (misspecified graph).}
In practice, the graph encoding the dependency structure among nodes may be only partially correct. To assess the robustness of GFDCM to graph misspecification, we generate the data exactly as in Scenario A (same distributional predictors, scalar covariates, coefficient functions $\beta_j(\cdot)$, censoring mechanism), but fit GFDCM with a supplied graph $\widetilde{\mathcal{E}}$ that differs from the true graph $\mathcal{E}$. Specifically, the true edge $(1,2)$ is omitted, and three spurious edges are added: $(2,3)$, which links two active nodes with different effects ($5\cos(2\pi p)$ and $3\sin(\pi p)$), and $(5,6)$ and $(1,8)$, which link active nodes to null nodes, so that
\[
\widetilde{\mathcal{E}} = \{(3,4), (3,5), (4,5), (6,7), (8,9), (8,10), (9,10), (2,3), (5,6), (1,8)\}.
\]
The edges among nodes $\{3,4,5\}$ are retained. We consider $n=300$ and $M=100$ Monte Carlo replications for this additional misspecified scenario.

\subsection{Simulation Results}
\paragraph{Scenario A (correctly specified graph).}

We evaluate the performance of the proposed GFDCM in terms of selection, fusion, and estimation accuracy. We compare the estimation performance of the proposed
GFDCM, which achieves graph-guided fusion and selection of node-level distributional representations with i) a functional Cox model \citep{gellar2015cox} using the 10 node-level distributional predictors $Q_{ij}(p)$ as functional covariates without any sparsity or fusion penalty, and ii) a groupLASSO version of the proposed objective function (8), which achieves node selection \citep{yue2025variable} , but does not introduce fusion between the nodes ($\lambda_2=0$). For the proposed GFDCM method, the tuning parameters are chosen using the proposed BIC criterion (21), for the other two methods, the tuning parameters are chosen based on the recommended criterion of each approach. The maximum number of ADMM iterations is set to 100 in the simulations. Note that only the GFDCM can achieve sparse, fused, and smooth estimates. We report the selection performance of GFDCM, across all sample sizes in Table \ref{tab1} which reports the selection percentages of each node (ROI) across the 100 Monte Carlo replications, along with the average true positive rate and false positive rate.

\begin{table}[ht]
\centering
\caption{Variable Selection Performance of the proposed GFDCM. We display the selection percentages ($\%$) of each node across all the Monte Carlo (M.C) replications. The average true positive rate (TPR), false positive rate (FPR) are reported below as a summary of the performance in percentages.}
\label{tab1}
\begin{tabular}{lccc}
\toprule
 & \multicolumn{3}{c}{Scenario} \\
\cmidrule(lr){2-4}
Node & \begin{tabular}[c]{@{}c@{}}Sample size\\ (n=300)\end{tabular} & \begin{tabular}[c]{@{}c@{}}Sample size\\ (n=500)\end{tabular} & \begin{tabular}[c]{@{}c@{}}Sample size \\ (n=1000)\end{tabular}\\
\midrule
1 & 100 & 100 & 100\\
2 & 100  & 100& 100 \\
3 & 100  & 100& 100\\
4 & 100 & 100 & 100\\
5 & 100  & 100& 100\\
\midrule
6 & 6 & 0& 0\\
7 & 7 & 2& 0\\
8 & 5& 2& 0\\
9 & 5  &2 & 0\\
10 & 5 & 2& 0\\
\midrule
TPR & 100 & 100& 100\\
FPR & 5.6 & 1.6& 0\\
\bottomrule
\end{tabular}
\end{table}

We observe that GFDCM exhibits a high true positive rate (selecting all the true nodes) and a negligible false positive rate, which declines as the sample size increases. Next, we report the fusion performance of GFDCM in terms of the selection percentages of the number of active groups identified in Table 2. Note that based on our data generation, there are two active groups (nodes $1-2$, and nodes $3-4-5$). It can be observed, across the three sample sizes, GFDCM exhibits a high fusion accuracy (ranging from $93\%-100\%$).

We report the estimation performance of GFDCM in terms of mean integrated squared error (MISE) of the distributional parameter estimates $\hat{\beta}_j(p)$, defined as
$$MISE_j = \frac{1}{M}\sum_{m=1}^{M} \left( \int_0^1 \left( \hat{\beta}_{j}(p) - \beta_j(p) \right)^2 \, dp \right).
$$
The MISE of the distributional effects from GFDCM are shown in Table \ref{tab:my-table3} along with the MISE of the estimates from the i) functional Cox model (FCM) and ii) groupLASSO (GLASSO).

\begin{table}[ht]
\centering
\caption{Fusion Performance of the proposed GFDCM. We display the percentages ($\%$) number of active groups of nodes identified across all the Monte Carlo (M.C) replications. The correct fusion (CF) percentage is reported below as a summary of the performance.}
\begin{tabular}{lccc}
\toprule
 & \multicolumn{3}{c}{Scenario} \\
\cmidrule(lr){2-4}
Unique Groups & \begin{tabular}[c]{@{}c@{}}Sample Size \\ (n=300)\end{tabular} & \begin{tabular}[c]{@{}c@{}}Sample Size \\ (n=500)\end{tabular} & \begin{tabular}[c]{@{}c@{}}Sample Size \\ (n=1000)\end{tabular}\\
\midrule
2 &  93 &98& 100\\
3 &   3 &2& 0\\
4 &  4 &0& 0\\
\bottomrule
CF &  93 &98& 100\\
\end{tabular}
\end{table}

\begin{table}[ht]
\centering
\caption{ Mean integrated squared error (MISE) of the active distributional effects ($\beta_j(p)$, $j=1,\ldots 5$) from the GFDCM across three sample sizes. The performance from FCM and GLASSO is reported in the parenthesis.}
\label{tab:my-table3}
\begin{tabular}{|l|l|l|l|l|l|}
\hline
Sample  Size  & $\beta_1(\cdot)$ & $\beta_2(\cdot)$ & $\beta_3(\cdot)$ & $\beta_4(\cdot)$ & $\beta_5$ \\ \hline

n=300               & 0.72 (1.08,2.66)     & 0.72(1.10,2.88)      & 0.41 (0.42,2.50)     & 0.41 (0.44,2.89)     & 0.41 (0.49,2.16)  \\

n=500               & 0.37 (0.63,1.57)     & 0.37(0.74,1.55)      & 0.22 (0.22,0.96)     & 0.22 (0.25,0.87)     & 0.22 (0.21,1.04)     \\
n=1000               & 0.19 (0.40,1.11)     & 0.19 (0.38,1.15)      & 0.13 (0.11,0.46)     & 0.13 (0.14,0.50)     & 0.13 (0.13,0.45)     \\ 

\hline
\end{tabular}
\end{table}

\begin{figure}[ht]
\begin{center}
\includegraphics[width=.8\linewidth , height=.6\linewidth]{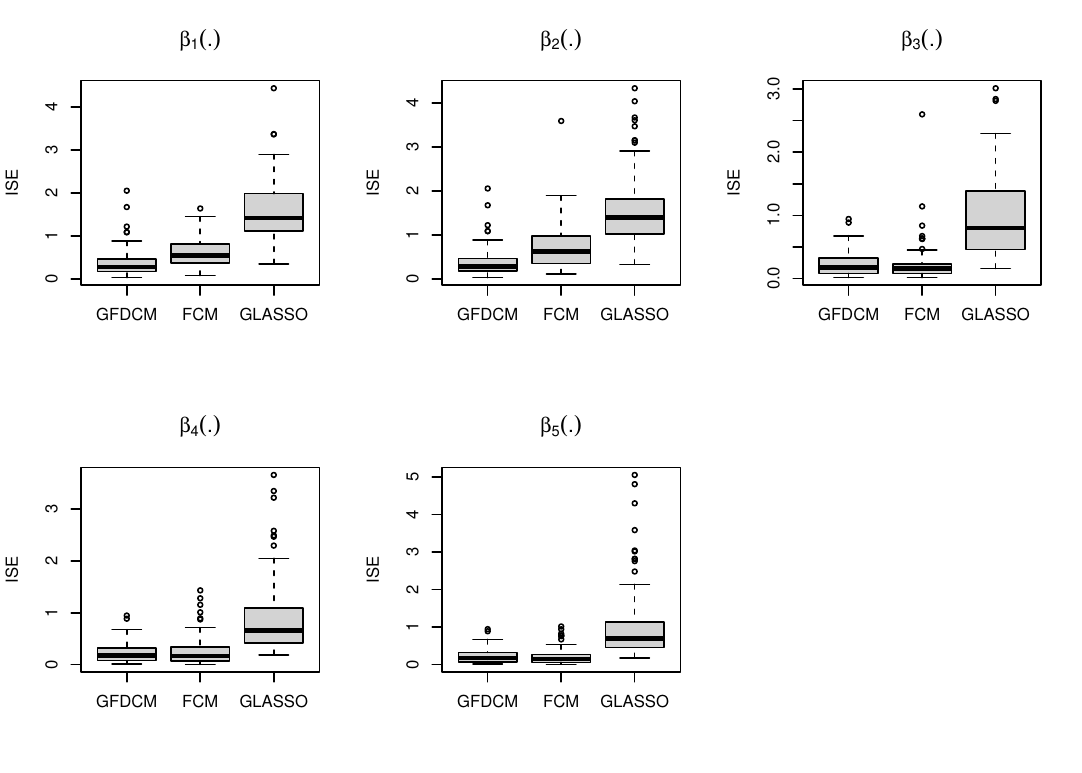}
\caption{The distribution of ISE loss for the three methods; GFDCM, FCM and GLASSO, sample size $n=500$.}
\label{fig:figsim1}
\end{center}
\end{figure}

\begin{figure}[ht]
\begin{center}

\includegraphics[width=.7\linewidth , height=.5\linewidth]{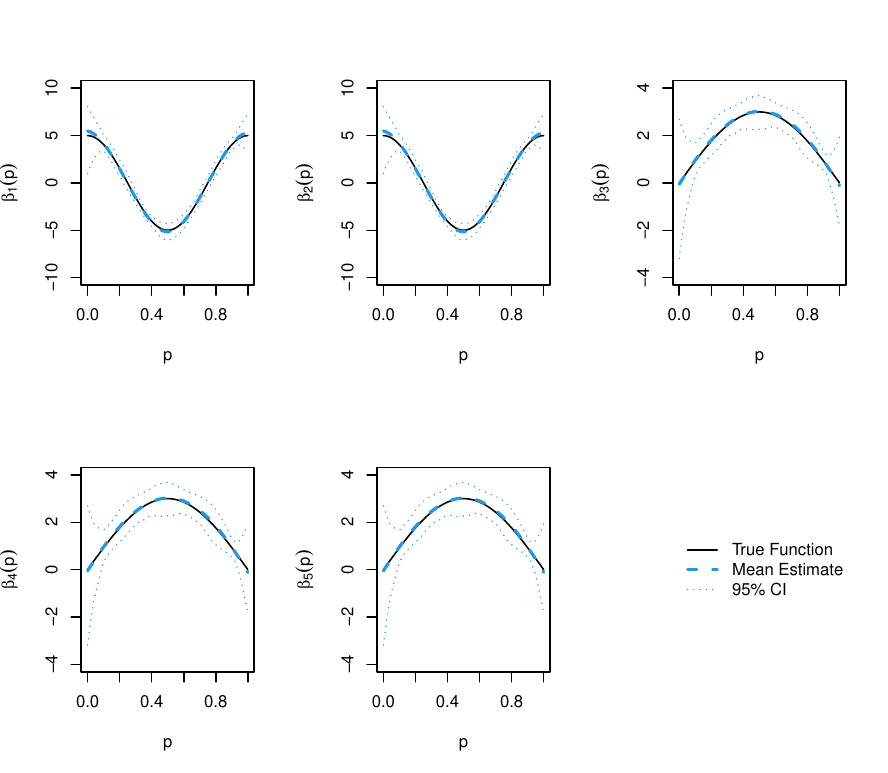}
\end{center}
\caption{Monte Carlo mean estimates from GFDCM and point-wise confidence intervals of the distributional coefficient functions $\beta_j(\cdot)$, for $n=500$.}
\label{fig:fig4}
\end{figure}

We can observe that accounting for the dependency structure via graph-guided fusion and imposing sparsity in our GFDCM method dramatically improves the estimation performance of the distributional effects, as compared to the FCM (no sparsity or fusion) and GLASSO (no fusion). The overall estimation performance improves with increasing sample size for all the methods. On average, the GFDCM estimators can be seen to provide $1.3$ times smaller MISE compared to FCM and 4.6 times smaller MISE compared to the GLASSO method, highlighting the benefit of the joint imposition of sparsity and fusion. We also display the distribution of ISE loss for the three methods in Figure \ref{fig:figsim1}, for a sample size of $n=500$, where an impressive performance of GFDCM can again be observed.
The Figures for $n=300,1000$ are shown in Web Figures 1 and 2, which exhibit a similar pattern.
We display the Monte Carlo mean estimate of the distributional effects $\beta_j(\cdot)$ from the GFDCM method, along with the percentile-based confidence intervals (across all replications) in Figure \ref{fig:fig4}, for sample size $n=500$. The GFDCM estimates are seen to capture the true distributional effects accurately, similar patterns are observed for sample sizes $n=300,1000$ and are reported in Web Figures 3 and 4.

\paragraph{Scenario B (misspecified graph).}
The detailed selection, fusion, and estimation results under the misspecified graph are reported in Web Appendix A and Web Tables 2--4. Overall, GFDCM selected all active nodes in every replicate and a misspecified graph did not lead to spurious fusion. However, little fusion was selected, and the false positive rate increased ($13.4\%$ vs $ 5.6\%$ under the correct graph). The estimation performance was substantially degraded relative to the correctly specified graph.

\section{ADNI Application}
\label{sec:4}

Our motivating application is based on a
neuroimaging dataset from the Alzheimer's Disease Neuroimaging
Initiative (ADNI).
In such data, the high-dimensional voxel-level measurements exhibit
rich dependency structures reflecting the brain's spatial and
anatomical dependencies.
Traditional scalar-on-image regression framework-based whole-brain
analyses \citep{reiss2010functional, wang2017generalized} are often
hard to interpret clinically.
Incorporating region-of-interest (ROI)-level information can therefore
aid in interpretability.
However, instead of reducing whole-brain image data to simple
ROI-level scalar summaries, we model the full distribution of voxel
intensities within each ROI to understand their association with time
to conversion to Alzheimer's disease.
Compared to whole-brain analysis \citep{kloppel2008automatic,
manning2017comparison}, which can be computationally intensive or
relies on scalar summaries, the ROI-level distributional
representations can provide additional interpretability by looking at
specific parts of the brain which contain the relevant prognostic
information, and by investigating distributional characteristics
beyond mean and traditional summaries. The main objective of our analysis is to illustrate the GFDCM
method by identifying brain regions and their corresponding
ROI-level distributional representations associated with time to
conversion to Alzheimer's disease.
Identifying brain regions predictive of progression to AD has
received considerable attention in neuroimaging research
\citep{zhang2015detection}, particularly within longitudinal cohort
studies such as ADNI. 

Our analytical cohort for the survival analysis includes $n=308$ subjects with available baseline MRI and longitudinal follow-up from the Alzheimer’s Disease Neuroimaging Initiative (ADNI), specifically the ADNI-1 cohort. 
We used only ADNI-1 to ensure a homogeneous cohort with consistent imaging protocols and adequate longitudinal follow-up for survival analysis. The subjects were mildly cognitively impaired at the baseline, and among them $184$ individuals progressed to AD, and $124$ were right-censored. The descriptive summaries of the sample are provided in Web Table 1. The survival outcome was defined as time from baseline imaging to AD diagnosis, measured in years, with observed follow-up ranging from approximately $0.48$ to $16.6$ years (median follow-up $2.54$ years). Baseline structural features were derived from T1-weighted MRI scans processed using the standardized ADNI preprocessing pipeline, including corrections for imaging artifacts and intensity inhomogeneity, followed by spatial normalization to a common template. Each brain was parcellated into $62$ anatomically defined ROIs using the DKT atlas, and voxel-level intensity values within each region were extracted and log-transformed to stabilize variability and remove possible skewness. Rather than summarizing these measurements using scalar metrics, we constructed the distributional representations $Q_{ij}(p)$ for each ROI ($j=1,\ldots,62$) using a fixed equispaced grid of $m=50$ quantile levels in $[0,1]$. In addition to these imaging-derived predictors, baseline demographic and clinical covariates, including age, sex, APOE4 status, were incorporated into the model. 

\begin{figure}[ht]
\begin{center}
\begin{tabular}{ll}
\includegraphics[width=.45\linewidth , height=.5\linewidth]{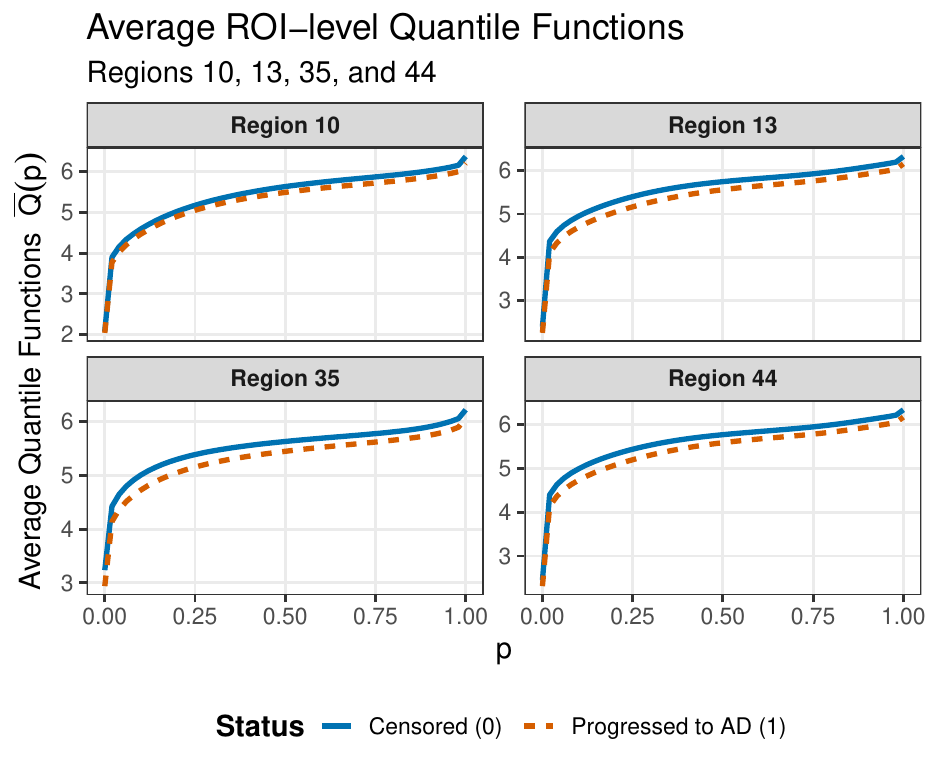} &
\includegraphics[width=.45\linewidth , height=.5\linewidth]{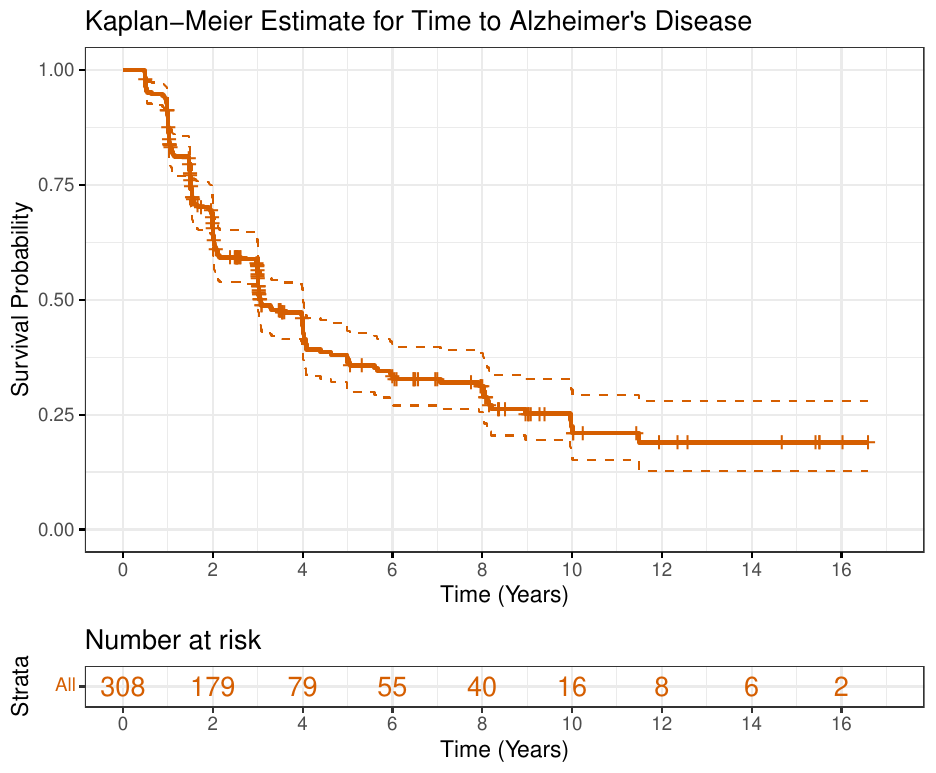}
\end{tabular}
\end{center}
\caption{Left: Displayed are the averages of ROI-level quantile functions of voxel-intensities (log-transformed) of 4 example regions (10,13,35,44) stratified by the AD status in the ADNI application. Right: KM estimate of survival probability (no progression to AD) in the ADNI application.}
\label{fig:fig3s2}
\end{figure}

Figure \ref{fig:fig3s2} (left panel) displays the average ROI-level quantile functions of log-transformed voxel intensities for selected representative regions, stratified by AD progression status. It can be observed that, the differences between subjects who progressed to AD and those who remained stable vary substantially across quantile levels and ROIs. The averages of ROI-level quantile functions of voxel-intensities (log-transformed) across all 62 regions are shown in Web Figure 5.
This empirical evidence motivates the use of distributional representations of ROI intensities within a survival modeling framework, as implemented in the proposed GFDCM approach. Figure \ref{fig:fig3s2} (right panel) displays the Kaplan-Meier estimate of the survival probability for time to Alzheimer’s disease (AD) conversion in our analytical cohort. The presence of both early events and long-term censoring can be observed among the individuals, indicating considerable heterogeneity in progression that may not be well captured using summary-based measures.


\begin{figure}[ht]
\begin{center}
\includegraphics[width=1\linewidth , height=.6\linewidth]{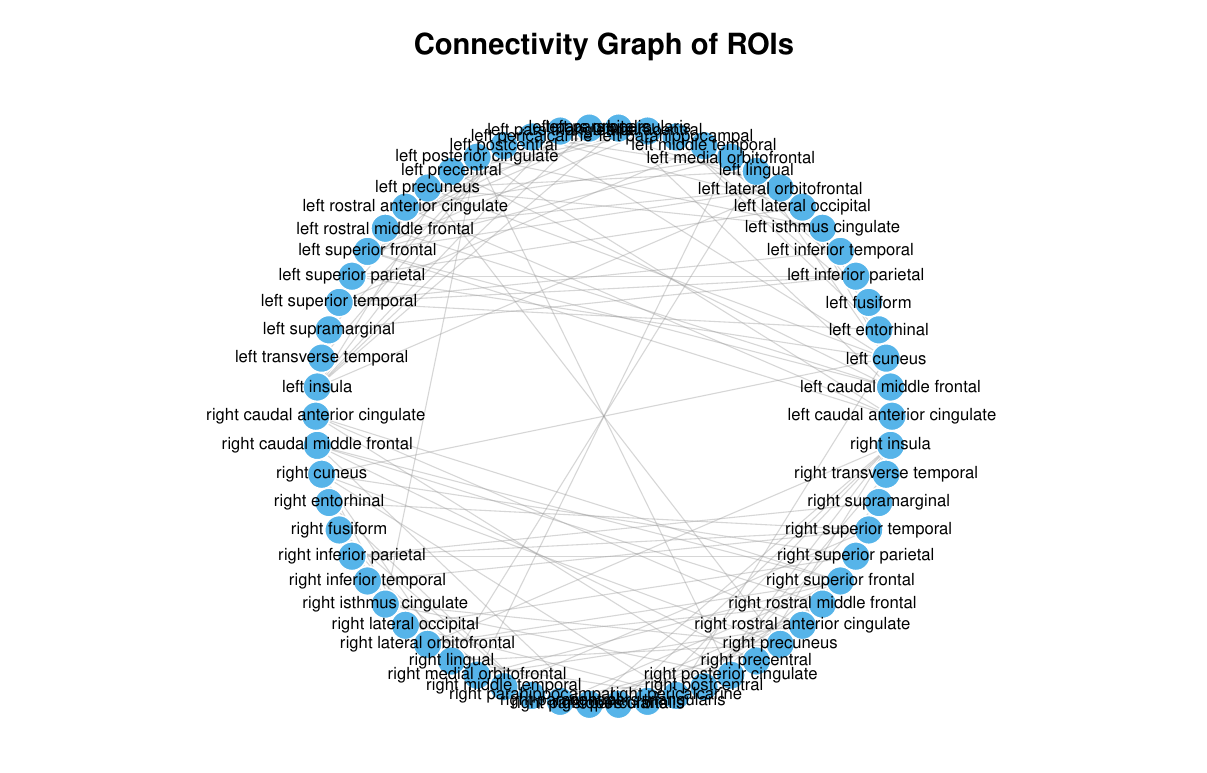}
\end{center}
\caption{The connectivity graph between the 62 regions (ROIs) in the Desikan-Killiany-Tourville (DKT) atlas corresponding to our ADNI application. Edges indicate connected regions are targeted for fusion in GFDCM.}
\label{fig:fig4s2}
\end{figure}

To incorporate anatomical structure into the modeling framework of GFDCM, we construct a connectivity graph over the $62$ ROIs defined by the DKT atlas, where edges represent spatial adjacency between regions based on shared anatomical boundaries. This graph, shown in Figure \ref{fig:fig4s2}, encodes prior structural relationships across neighboring brain regions and serves as the basis for the graph-fused regularization in GFDCM by providing the dependency structure between the ROIs via an undirected graph $G=(V,E)$. 


We consider the following distributional Cox model to model the hazard of AD at time $t$ as,
\begin{equation}
\label{eq:cox_model}
\lambda_i(t \mid \*Z_i, Q_{ij}(\cdot),j=1,\ldots,J) = \lambda_0(t) \exp\left( \*Z_i^T\bm\alpha+ \sum_{j=1}^J \int_0^1 Q_{ij}(p) \beta_j(p) \, dp \right),
\end{equation}
where $\*Z_i$ constitutes the scalar covariates age, sex (sex=1 for male, 0 for female), APOE4 status for subject $i$, and $Q_{ij}(p)$ are the ROI-level distributional representations of voxel intensities at baseline. We apply the proposed GFDCM with the known dependency structure $G=(V,E)$, to identify the influential ROI-level distributional predictors and quantify their association with the hazard of AD. The tuning parameters are selected based on the proposed BIC criterion in Section 2.6, and $d=5$ basis functions were used as the optimal number of basis for modelling the distributional coefficient functions. The maximum number of ADMM iterations is set to 500 for this case study.

The proposed GFDCM identifies the ROI 44: ``right middle temporal", as the sole region whose distributional representation is associated with time to Alzheimer’s disease. 

\begin{figure}[h!]
\begin{center}
\includegraphics[width=0.55\linewidth , height=0.5\linewidth]{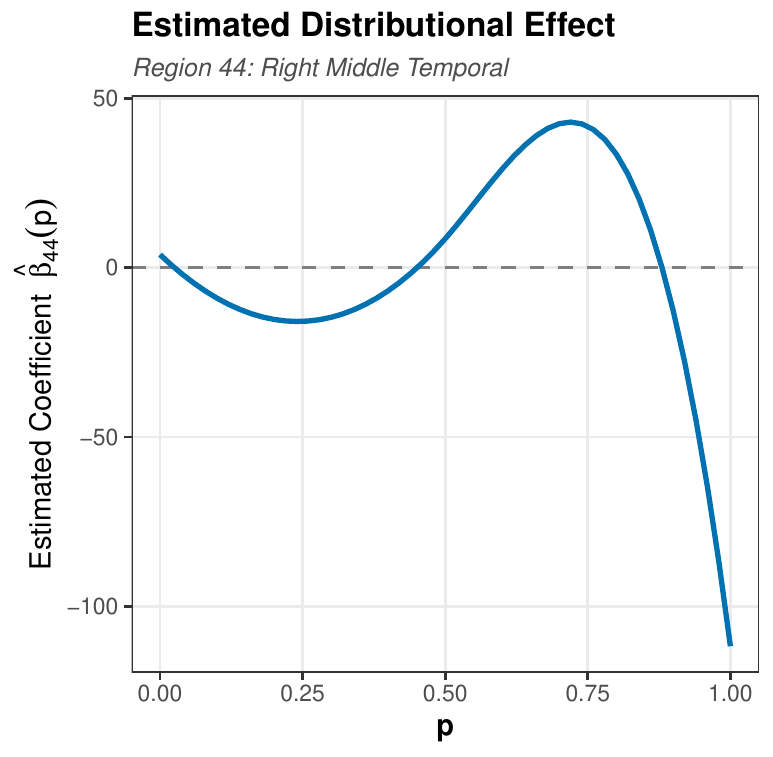}
\end{center}
\caption{The estimated distributional effect $\hat{\beta}_{44}(p)$ of the selected Region 44 (Right Middle Temporal) on log-hazard of progression to AD based on the GFDCM in the ADNI application. }
\label{fig:fig5s2}
\end{figure}

Figure \ref{fig:fig5s2} presents the estimated distributional effect $\hat{\beta}_{44}(p)$ for Region 44 (right middle temporal) as a function of the quantile level $p$. The estimated coefficient function exhibits pronounced nonlinearity and heterogeneity across the quantile levels, and can be interpreted as capturing a continuous contrast across ROI-level intensity quantiles. Specifically, the effect captures a contrast across quantiles: $p\in (0.48,0.85)$-getting positive weights, quantiles $p\in (0.85,1)$-getting high negative weights, quantiles $p\in (0,0.48)$-getting small negative weights. Hence, a higher maximal quantile $p\in (0.85,1)$ of voxel intensity in this region, would be associated with a reduced hazard of AD. Interestingly, near median ($p=0.5$), the effect size is small, illustrating the importance of looking at distributional representations and maximal quantiles of ROI-level voxel intensities, beyond scalar summaries such as the mean intensity. Such heterogeneous, quantile-specific effects would not be captured by models based on scalar summaries.


The robustness of the GFDCM method is further highlighted by the evaluation of candidate models along the regularization path. Based on the proposed BIC criterion and GFDCM, the second best model (yielding the second smallest BIC) identified the ROI 13: ``left middle temporal", as the sole region whose distributional representation is associated with time to Alzheimer’s disease.  
The estimated distributional effect $\beta_{13}(p)$ is displayed in Web Figure 6 and can be seen to capture a similar contrast between the voxel-intensity quantiles as ROI 44. These two regions (ROI 44: right middle temporal and ROI 13: left middle temporal) have been widely reported in AD-related neurodegeneration \citep{karas2008amnestic,chen2022spatially}, lending biological plausibility to the selected models. 
The selection of only a small subset of regions highlights the effectiveness of the proposed GFDCM framework in performing structured variable selection in a high-dimensional setting with a limited sample size, while leveraging both distributional information and spatial dependence.


\begin{figure}[h!]
\begin{center}
\includegraphics[width=0.8\linewidth , height=0.6\linewidth]{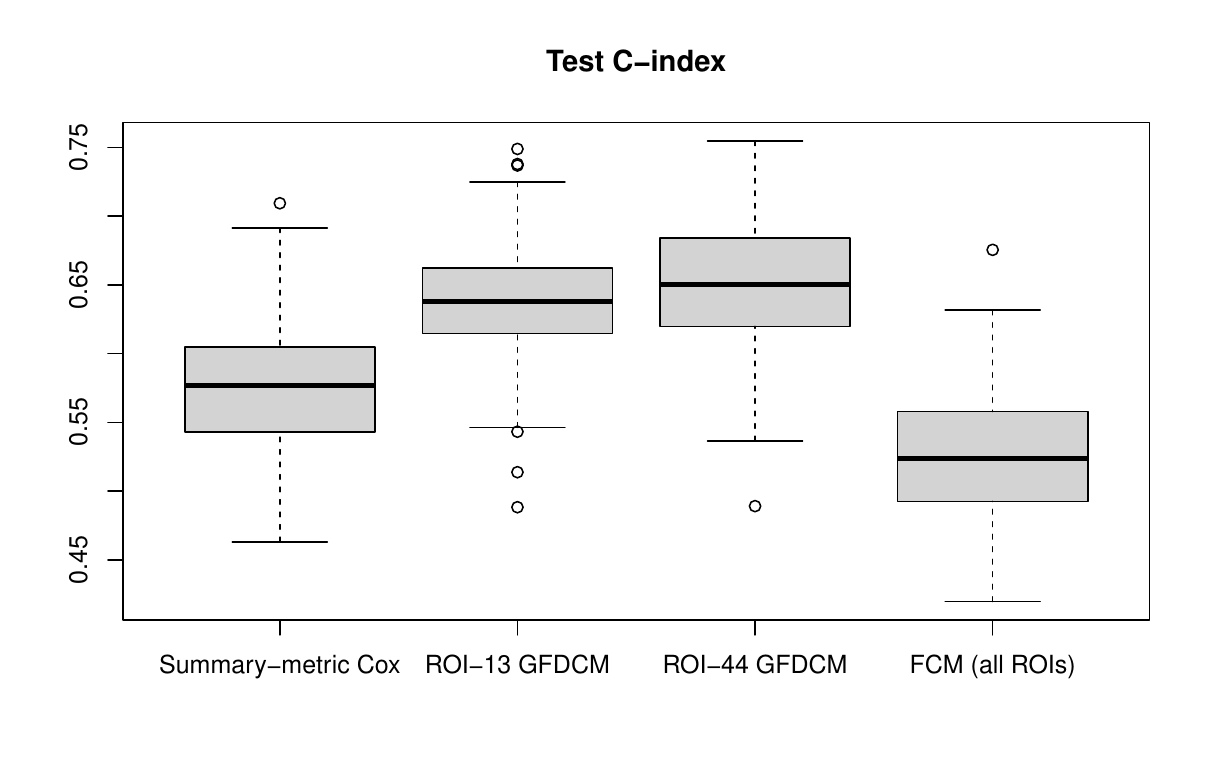}
\end{center}
\caption{The average test set C-Index (across 100 repeated train-test) from i) a summary metric-based Cox model, using the mean of 62 ROIs, ii) GFDCM using ROI 13, iii) GFDCM using ROI 44 iv) FCM using all ROIs. The ROI-44 (Right Middle Temporal) based GFDCM, and ROI-13 (Left Middle Temporal) based GFDCM are the top 2 selected models using the proposed method, when ranked via BIC.}
\label{fig:fig6s2}
\end{figure}

Finally, we evaluate the out-of-sample predictive performance of the proposed GFDCM framework using repeated train-test splits, and compare the results to a summary-metric-based Cox model using the mean intensity of the 62 ROIs as predictors in a multivariable Cox model. Note that an unpenalized FCM is not even estimable in this case with 62 regions, since $62\times 5+3=313$ leads to a higher number of parameters than $n=308$ and events (184), and hence regularization is essential.  Even with a smoothing penalty, the second-order roughness penalty leaves the linear part of each coefficient function unpenalized, so $3+62\times 2=127$ parameters remain unpenalized, too many to estimate reliably from $184$ events. For completeness, we nonetheless include this smoothing-penalized FCM with all 62 ROIs as a competitor. Because estimating a separate smoothing parameter for each of the 62 coefficient functions by REML was computationally prohibitive, we use a single smoothing parameter shared across all ROIs, analogous to the common smoothness parameter $\psi$ in GFDCM, and select it by BIC from a grid ($10^{-1},10^{0},\ldots,10^{5}$) using the training data only in each split.

Predictive accuracy is assessed using the concordance index (C-index) \citep{harrell1996multivariable}. The results are displayed in Figure \ref{fig:fig6s2}. The GFDCM model using Region 44 (Right Middle Temporal), selected based on the smallest BIC, achieves a $13.1\%$ improvement in average C-index (0.65) relative to the summary metric-based Cox model (0.57), and a $22.8\%$ improvement in average C-index relative to the FCM using all ROIs (0.53). This illustrates that, without sparsity, the FCM can be largely overfit in such a small sample. Similarly, the model based on Region 13 (Left Middle Temporal) yields an $11.4\%$ improvement in predictive performance based on the average C-Index (0.64) compared to the summary-based model. These results demonstrate that incorporating ROI-based distributional information has substantial potential to enhance the ability to predict progression to Alzheimer’s disease and underscores the limitations of traditional models that rely solely on mean-based summaries of imaging data.

\section{Discussion}
\label{disc}
In this article, we have developed the Graph-Fused Distributional Cox
Model (GFDCM), a penalized Cox regression framework with distribution-valued predictors organized by a known graph. Instead of using the node-level summaries of measurements, GFDCM considers the full distribution of node-level measurements as predictors, retaining node-specific interpretability while capturing more information than summary-based measures. The proposed method takes into account the known spatial relationships among nodes through graph-based fusion, allowing information sharing across anatomically adjacent nodes, and achieves simultaneous selection and fusion of nodes while estimating their distributional effects across varying quantile levels. The simulation results illustrate satisfactory selection, fusion, and estimation accuracy of GFDCM.  The application of GFDCM to ROI-based distributional
representations of baseline T1-weighted MRI data collected during ADNI-1 identified highly interpretable brain regions associated with AD conversion, along with the distributional effects of ROI-level voxel intensity distribution. The GFDCM also provided superior predictive performance compared to the traditional Cox model using ROI-based summary metrics. Hence, the GFDCM can be used for biomarker development with ROI-based distributional representations and early screening for AD.

The proposed GFDCM method can be extended in multiple directions. In the current article, we considered linear effects of the distributional representations; nonlinear and additive effects would be interesting to explore in future \citep{cui2021additive}, which might provide significant predictive gains. We have also considered a proportional hazard framework in the proposed GFDCM. It would be interesting to extend the method to more flexible survival models, such as the AFT, functional time-transformation model \citep{ghosal2025functional}. Finally, GFDCM uses the baseline ROI-level distributional representations. In the future, it would be of interest to integrate longitudinal imaging modalities into GFDCM through a joint modelling framework, which can provide additional prognostic insight into disease progression \citep{ghosh2025ensemble}, and remain an area of future research based on our current proposal.

\section*{Supplementary Material}
Web Appendix A, Web Tables 1-4, and Web Figures 1--6 referenced in Sections \ref{sec:3} and \ref{sec:4} are available with this paper at the Biostatistics website.

\section*{Disclosure Statement}
The authors report there are no competing interests to declare.

\section*{Software}
Software illustration of the proposed method is provided with this paper and will be made available online at Github.

\section*{Data Availability Statement}
Data used in the preparation of this article were obtained from the Alzheimer’s Disease Neuroimaging Initiative (ADNI) database (\url{http://adni.loni.usc.edu/}). Investigators within ADNI contributed to the design and implementation of ADNI and/or provided data, but did not participate in the analysis or writing of this report. 
A complete listing of ADNI investigators can be found at: \url{http://adni.loni.usc.edu/wp-content/uploads/how_to_apply/ADNI_Acknowledgement_List.pdf}.

\singlespacing
\bibliographystyle{biorefs}
\bibliography{ref,ref_DJ}

@article{petersen2010alzheimer,
  title={Alzheimer's disease Neuroimaging Initiative (ADNI) clinical characterization},
  author={Petersen, Ronald Carl and Aisen, Paul S and Beckett, Laurel A and Donohue, Michael C and Gamst, Anthony Collins and Harvey, Danielle J and Jack Jr, Clifford R and Jagust, William J and Shaw, Leslie M and Toga, Arthur W and others},
  journal={Neurology},
  volume={74},
  number={3},
  pages={201--209},
  year={2010},
  publisher={Lippincott Williams \& Wilkins}
}

@article{chen2008extended,
  title={Extended Bayesian information criteria for model selection with large model spaces},
  author={Chen, Jiahua and Chen, Zehua},
  journal={Biometrika},
  volume={95},
  number={3},
  pages={759--771},
  year={2008},
  publisher={Oxford University Press}
}

@article{wang2017generalized,
  title={Generalized scalar-on-image regression models via total variation},
  author={Wang, Xiao and Zhu, Hongtu and Alzheimer’s Disease Neuroimaging Initiative},
  journal={Journal of the American Statistical Association},
  volume={112},
  number={519},
  pages={1156--1168},
  year={2017},
  publisher={Taylor \& Francis}
}

@article{reiss2010functional,
  title={Functional generalized linear models with images as predictors},
  author={Reiss, Philip T and Ogden, R Todd},
  journal={Biometrics},
  volume={66},
  number={1},
  pages={61--69},
  year={2010},
  publisher={Oxford University Press}
}

@article{hyndman1996sample,
  title={Sample quantiles in statistical packages},
  author={Hyndman, Rob J and Fan, Yanan},
  journal={The American Statistician},
  volume={50},
  number={4},
  pages={361--365},
  year={1996},
  publisher={Taylor \& Francis}
}

@article{ghosal2021scalar,
  title={Scalar on time-by-distribution regression and its application for modelling associations between daily-living physical activity and cognitive functions in Alzheimer’s Disease},
  author={Ghosal, Rahul and Varma, Vijay R and Volfson, Dmitri and Urbanek, Jacek and Hausdorff, Jeffrey M and Watts, Amber and Zipunnikov, Vadim},
  journal={Scientific Reports},
  volume={12},
  number={1},
  pages={1--16},
  year={2022},
  publisher={Nature Publishing Group}
}

@article{petersen2019frechet,
  title={Fr{\'e}chet regression for random objects with Euclidean predictors},
  author={Petersen, Alexander and M{\"u}ller, Hans-Georg},
journal={The Annals of Statistics},
volume={47},
  number={2},
  pages={691--719},
  year={2019}
}

@article{yang2020random,
  title={Random distributional response model based on spline method},
  author={Yang, Hojin},
  journal={Journal of Statistical Planning and Inference},
  volume={207},
  pages={27--44},
  year={2020},
  publisher={Elsevier}
}

@article{bhattacharjee2023single,
  title={Single index Fr{\'e}chet regression},
  author={Bhattacharjee, Satarupa and M{\"u}ller, Hans-Georg},
  journal={The Annals of Statistics},
  volume={51},
  number={4},
  pages={1770--1798},
  year={2023},
  publisher={Institute of Mathematical Statistics}
}

@article{matabuena2025predicting,
  title={Predicting distributions of physical activity profiles in the National Health and Nutrition Examination Survey database using a partially linear Fr{\'e}chet single index model},
  author={Matabuena, Marcos and Ghosal, Aritra and Meiring, Wendy and Petersen, Alexander},
  journal={Biostatistics},
  volume={26},
  number={1},
  pages={kxaf013},
  year={2025},
  publisher={Oxford University Press}
}

@article{bhattacharjee2025geodesic,
  title={Geodesic mixed effects models for repeatedly observed/longitudinal random objects},
  author={Bhattacharjee, Satarupa and M{\"u}ller, Hans-Georg},
  journal={Journal of the American Statistical Association},
  pages={1--23},
  year={2025},
  publisher={Taylor \& Francis}
}

@article{ghosh2025ensemble,
  title={Ensemble survival analysis for preclinical cognitive decline prediction in Alzheimer's disease using longitudinal biomarkers},
  author={Ghosh, Dhrubajyoti and Pal, Samhita and Lutz, Michael and Luo, Sheng and Alzheimer’s Disease Neuroimaging Initiative},
  journal={Journal of Alzheimer’s Disease},
  volume={107},
  number={3},
  pages={1256--1266},
  year={2025},
  publisher={SAGE Publications Sage UK: London, England}
}

@article{kloppel2008automatic,
  title={Automatic classification of MR scans in Alzheimer's disease},
  author={Kl{\"o}ppel, Stefan and Stonnington, Cynthia M and Chu, Carlton and Draganski, Bogdan and Scahill, Rachael I and Rohrer, Jonathan D and Fox, Nick C and Jack Jr, Clifford R and Ashburner, John and Frackowiak, Richard SJ},
  journal={Brain},
  volume={131},
  number={3},
  pages={681--689},
  year={2008},
  publisher={Oxford University Press}
}

@misc{who2025dementia,
  title={Dementia: Key Facts},
  author={{World Health Organization}},
  year={2025},
  url={https://www.who.int/news-room/fact-sheets/detail/dementia},
  note={Accessed: 2026-04-13}
}

@article{hebert2013alzheimer,
  title={Alzheimer disease in the United States (2010--2050) estimated using the 2010 census},
  author={Hebert, Liesi E and Weuve, Jennifer and Scherr, Paul A and Evans, Denis A},
  journal={Neurology},
  volume={80},
  number={19},
  pages={1778--1783},
  year={2013},
  publisher={AAN Enterprises}
}

@article{yu2022mapping,
  title={Mapping the genetic-imaging-clinical pathway with applications to alzheimer’s disease},
  author={Yu, Dengdeng and Wang, Linbo and Kong, Dehan and Zhu, Hongtu},
  journal={Journal of the American Statistical Association},
  volume={117},
  number={540},
  pages={1656--1668},
  year={2022},
  publisher={Taylor \& Francis}
}

@article{lee2015bflcrm,
  title={BFLCRM: A Bayesian functional linear Cox regression model for predicting time to conversion to Alzheimer’s disease},
  author={Lee, Eunjee and Zhu, Hongtu and Kong, Dehan and Wang, Yalin and Giovanello, Kelly Sullivan and Ibrahim, Joseph G},
  journal={The annals of applied statistics},
  volume={9},
  number={4},
  pages={2153},
  year={2015}
}

@article{yang2021weighted,
  title={Weighted functional linear Cox regression model},
  author={Yang, Hojin and Zhu, Hongtu and Ahn, Mihye and Ibrahim, Joseph G},
  journal={Statistical Methods in Medical Research},
  volume={30},
  number={8},
  pages={1917--1931},
  year={2021},
  publisher={SAGE Publications Sage UK: London, England}
}

@article{kong2018flcrm,
  title={FLCRM: Functional linear cox regression model},
  author={Kong, Dehan and Ibrahim, Joseph G and Lee, Eunjee and Zhu, Hongtu},
  journal={Biometrics},
  volume={74},
  number={1},
  pages={109--117},
  year={2018},
  publisher={Wiley Online Library}
}

@article{manning2017comparison,
  title={A comparison of accelerated and non-accelerated MRI scans for brain volume and boundary shift integral measures of volume change: evidence from the ADNI dataset},
  author={Manning, Emily N and Leung, Kelvin K and Nicholas, Jennifer M and Malone, Ian B and Cardoso, M Jorge and Schott, Jonathan M and Fox, Nick C and Barnes, Josephine and Alzheimer’s Disease Neuroimaging Initiative},
  journal={Neuroinformatics},
  volume={15},
  number={2},
  pages={215--226},
  year={2017},
  publisher={Springer}
}

@article{ghosal2025functional,
  title={Functional time transformation model with applications to digital health},
  author={Ghosal, Rahul and Matabuena, Marcos and Ghosh, Sujit K},
  journal={Computational Statistics \& Data Analysis},
  pages={108131},
  year={2025},
  publisher={Elsevier}
}

@article{karas2008amnestic,
  title={Amnestic mild cognitive impairment: structural MR imaging findings predictive of conversion to Alzheimer disease},
  author={Karas, G and Sluimer, J and Goekoop, R and Van Der Flier, W and Rombouts, SARB and Vrenken, H and Scheltens, P and Fox, N and Barkhof, F},
  journal={American Journal of Neuroradiology},
  volume={29},
  number={5},
  pages={944--949},
  year={2008},
  publisher={American Journal of Neuroradiology}
}

@article{harrell1996multivariable,
  title={Multivariable prognostic models: issues in developing models, evaluating assumptions and adequacy, and measuring and reducing errors},
  author={Harrell Jr, Frank E and Lee, Kerry L and Mark, Daniel B},
  journal={Statistics in medicine},
  volume={15},
  number={4},
  pages={361--387},
  year={1996},
  publisher={Wiley Online Library}
}

@article{chen2025high,
  title={High-dimensional partially linear functional Cox models},
  author={Chen, Xin and Liu, Hua and Men, Jiaqi and You, Jinhong},
  journal={Biometrics},
  volume={81},
  number={1},
  pages={ujae164},
  year={2025},
  publisher={Oxford University Press}
}

@article{matabuena2024multilevel,
  title={Multilevel functional distributional models with application to continuous glucose monitoring in diabetes clinical trials},
  author={Matabuena, Marcos and Crainiceanu, Ciprian M},
  journal={arXiv preprint arXiv:2403.10514},
  year={2024}
}

@article{chen2021wasserstein,
  title={Wasserstein regression},
  author={Chen, Yaqing and Lin, Zhenhua and M{\"u}ller, Hans-Georg},
  journal={Journal of the American Statistical Association},
  volume={118},
  number={542},
  pages={869--882},
  year={2023},
  publisher={Taylor \& Francis}
}

@article{matabuena2020glucodensities,
  title={Glucodensities: a new representation of glucose profiles using distributional data analysis},
  author={Matabuena, Marcos and Petersen, Alexander and Vidal, Juan C and Gude, Francisco},
  journal={Statistical Methods in Medical Research},
  volume={30},
  number={6},
  pages={1445-1464},
  year={2021}
}

@article{zhang2024scalar,
  title={A scalar-on-quantile-function approach for estimating short-term health effects of environmental exposures},
  author={Zhang, Yuzi and Chang, Howard H and Warren, Joshua L and Ebelt, Stefanie T},
  journal={Biometrics},
  volume={80},
  number={1},
  pages={ujae008},
  year={2024},
  publisher={Oxford University Press}
}

@article{ghosal2025distributional,
  title={Distributional outcome regression via quantile functions and its application to modelling continuously monitored heart rate and physical activity},
  author={Ghosal, Rahul and Ghosh, Sujit K and Schrack, Jennifer A and Zipunnikov, Vadim},
  journal={Journal of the American Statistical Association},
  pages={1--13},
  year={2025},
  publisher={Taylor \& Francis}
}

@article{beer2019incorporating,
  title={Incorporating prior information with fused sparse group lasso: Application to prediction of clinical measures from neuroimages},
  author={Beer, Joanne C and Aizenstein, Howard J and Anderson, Stewart J and Krafty, Robert T},
  journal={Biometrics},
  volume={75},
  number={4},
  pages={1299--1309},
  year={2019},
  publisher={Oxford University Press}
}

@article{boyd2011distributed,
  title={Distributed optimization and statistical learning via the alternating direction method of multipliers},
  author={Boyd, Stephen and Parikh, Neal and Chu, Eric and Peleato, Borja and Eckstein, Jonathan and others},
  journal={Foundations and Trends{\textregistered} in Machine learning},
  volume={3},
  number={1},
  pages={1--122},
  year={2011},
  publisher={Now Publishers, Inc.}
}

@article{yuan2006model,
  title={Model selection and estimation in regression with grouped variables},
  author={Yuan, Ming and Lin, Yi},
  journal={Journal of the Royal Statistical Society Series B: Statistical Methodology},
  volume={68},
  number={1},
  pages={49--67},
  year={2006},
  publisher={Oxford University Press}
}

@article{ghosal2020variable,
  title={Variable selection in functional linear concurrent regression},
  author={Ghosal, Rahul and Maity, Arnab and Clark, Timothy and Longo, Stefano B},
  journal={Journal of the Royal Statistical Society Series C: Applied Statistics},
  volume={69},
  number={3},
  pages={565--587},
  year={2020},
  publisher={Oxford University Press}
}

@article{ranciati2021fused,
  title={Fused graphical lasso for brain networks with symmetries},
  author={Ranciati, Saverio and Roverato, Alberto and Luati, Alessandra},
  journal={Journal of the Royal Statistical Society Series C: Applied Statistics},
  volume={70},
  number={5},
  pages={1299--1322},
  year={2021},
  publisher={Oxford University Press}
}

@article{cuingnet2012spatial,
  title={Spatial and anatomical regularization of SVM: a general framework for neuroimaging data},
  author={Cuingnet, R{\'e}mi and Glaun{\`e}s, Joan Alexis and Chupin, Marie and Benali, Habib and Colliot, Olivier},
  journal={IEEE transactions on pattern analysis and machine intelligence},
  volume={35},
  number={3},
  pages={682--696},
  year={2012},
  publisher={IEEE}
}

@article{yue2025variable,
  author = {Yue, Yuanzhen and Self, Stella and Wu, Yichao and Zhang, Jiajia and Ghosal, Rahul},
    title = {Variable selection in functional linear Cox model},
    journal = {Biometrics},
    volume = {82},
    number = {1},
    pages = {ujag044},
    year = {2026},
    month = {03},
    issn = {0006-341X},
    doi = {10.1093/biomtc/ujag044},
    url = {https://doi.org/10.1093/biomtc/ujag044}
}

@article{sabuncu2011dynamics,
  title={The dynamics of cortical and hippocampal atrophy in Alzheimer disease},
  author={Sabuncu, Mert R and Desikan, Rahul S and Sepulcre, Jorge and Yeo, Boon Thye T and Liu, Hesheng and Schmansky, Nicholas J and Reuter, Martin and Weiner, Michael W and Buckner, Randy L and Sperling, Reisa A and others},
  journal={Archives of neurology},
  volume={68},
  number={8},
  pages={1040--1048},
  year={2011},
  publisher={American Medical Association}
}

@article{tibshirani2005sparsity,
  title={Sparsity and smoothness via the fused lasso},
  author={Tibshirani, Robert and Saunders, Michael and Rosset, Saharon and Zhu, Ji and Knight, Keith},
  journal={Journal of the Royal Statistical Society Series B: Statistical Methodology},
  volume={67},
  number={1},
  pages={91--108},
  year={2005},
  publisher={Oxford University Press}
}

@article{ghosal2023distributional,
  title={Distributional data analysis via quantile functions and its application to modeling digital biomarkers of gait in Alzheimer’s disease},
  author={Ghosal, Rahul and Varma, Vijay R and Volfson, Dmitri and Hillel, Inbar and Urbanek, Jacek and Hausdorff, Jeffrey M and Watts, Amber and Zipunnikov, Vadim},
  journal={Biostatistics},
  volume={24},
  number={3},
  pages={539--561},
  year={2023},
  publisher={Oxford University Press}
}

@article{yang2020quantile,
  title={Quantile function on scalar regression analysis for distributional data},
  author={Yang, Hojin and Baladandayuthapani, Veerabhadran and Rao, Arvind UK and Morris, Jeffrey S},
  journal={Journal of the American Statistical Association},
  volume={115},
  number={529},
  pages={90--106},
  year={2020},
  publisher={Taylor \& Francis}
}

@article{mueller2005ways,
  title={Ways toward an early diagnosis in Alzheimer’s disease: the Alzheimer’s Disease Neuroimaging Initiative (ADNI)},
  author={Mueller, Susanne G and Weiner, Michael W and Thal, Leon J and Petersen, Ronald C and Jack, Clifford R and Jagust, William and Trojanowski, John Q and Toga, Arthur W and Beckett, Laurel},
  journal={Alzheimer's \& Dementia},
  volume={1},
  number={1},
  pages={55--66},
  year={2005},
  publisher={Elsevier}
}

@article{jack2010hypothetical,
  title={Hypothetical model of dynamic biomarkers of the Alzheimer's pathological cascade},
  author={Jack, Clifford R and Knopman, David S and Jagust, William J and Shaw, Leslie M and Aisen, Paul S and Weiner, Michael W and Petersen, Ronald C and Trojanowski, John Q},
  journal={The Lancet Neurology},
  volume={9},
  number={1},
  pages={119--128},
  year={2010},
  publisher={Elsevier}
}

@article{weiner2017alzheimer,
  title={The Alzheimer's Disease Neuroimaging Initiative 3: Continued innovation for clinical trial improvement},
  author={Weiner, Michael W and Veitch, Dallas P and Aisen, Paul S and Beckett, Laurel A and Cairns, Nigel J and Green, Robert C and Harvey, Danielle and Jack Jr, Clifford R and Jagust, William and Morris, John C and others},
  journal={Alzheimer's \& Dementia},
  volume={13},
  number={5},
  pages={561--571},
  year={2017},
  publisher={Elsevier}
}

@article{jack2018nia,
  title={NIA-AA research framework: toward a biological definition of Alzheimer's disease},
  author={Jack Jr, Clifford R and Bennett, David A and Blennow, Kaj and Carrillo, Maria C and Dunn, Billy and Haeberlein, Samantha Budd and Holtzman, David M and Jagust, William and Jessen, Frank and Karlawish, Jason and others},
  journal={Alzheimer's \& dementia},
  volume={14},
  number={4},
  pages={535--562},
  year={2018},
  publisher={Wiley Online Library}
}

@article{petersen2016functional,
  title={Functional data analysis for density functions by transformation to a Hilbert space},
  author={Petersen, Alexander and M{\"u}ller, Hans-Georg},
  year={2016}
}

@article{tibshirani1997lasso,
  title={The lasso method for variable selection in the Cox model},
  author={Tibshirani, Robert},
  journal={Statistics in medicine},
  volume={16},
  number={4},
  pages={385--395},
  year={1997},
  publisher={Wiley Online Library}
}

@article{gellar2015cox,
  title={Cox regression models with functional covariates for survival data},
  author={Gellar, Jonathan E and Colantuoni, Elizabeth and Needham, Dale M and Crainiceanu, Ciprian M},
  journal={Statistical modelling},
  volume={15},
  number={3},
  pages={256--278},
  year={2015},
  publisher={SAGE Publications Sage India: New Delhi, India}
}

@article{zhu2017novel,
  title={A novel relational regularization feature selection method for joint regression and classification in AD diagnosis},
  author={Zhu, Xiaofeng and Suk, Heung-Il and Wang, Li and Lee, Seong-Whan and Shen, Dinggang and Alzheimer’s Disease Neuroimaging Initiative and others},
  journal={Medical image analysis},
  volume={38},
  pages={205--214},
  year={2017},
  publisher={Elsevier}
}

@article{ahmadzadeh2023neuroimaging,
  title={Neuroimaging and machine learning for studying the pathways from mild cognitive impairment to alzheimer’s disease: a systematic review},
  author={Ahmadzadeh, Maryam and Christie, Gregory J and Cosco, Theodore D and Arab, Ali and Mansouri, Mehrdad and Wagner, Kevin R and DiPaola, Steve and Moreno, Sylvain},
  journal={BMC neurology},
  volume={23},
  number={1},
  pages={309},
  year={2023},
  publisher={Springer}
}

@article{williams202112,
  title={12-year prediction of mild cognitive impairment aided by Alzheimer’s brain signatures at mean age 56},
  author={Williams, McKenna E and Elman, Jeremy A and McEvoy, Linda K and Andreassen, Ole A and Dale, Anders M and Eglit, Graham ML and Eyler, Lisa T and Fennema-Notestine, Christine and Franz, Carol E and Gillespie, Nathan A and others},
  journal={Brain Communications},
  volume={3},
  number={3},
  pages={fcab167},
  year={2021},
  publisher={Oxford University Press}
}

@article{petersen2022modeling,
  title={Modeling probability density functions as data objects},
  author={Petersen, Alexander and Zhang, Chao and Kokoszka, Piotr},
  journal={Econometrics and Statistics},
  volume={21},
  pages={159--178},
  year={2022},
  publisher={Elsevier}
}

@article{talska2021compositional,
  title={Compositional scalar-on-function regression with application to sediment particle size distributions},
  author={Talsk{\'a}, Ren{\'a}ta and Hron, Karel and Grygar, Tom{\'a}{\v{s}} Matys},
  journal={Mathematical Geosciences},
  volume={53},
  number={7},
  pages={1667--1695},
  year={2021},
  publisher={Springer}
}

@article{matabuena2023distributional,
  title={Distributional data analysis of accelerometer data from the NHANES database using nonparametric survey regression models},
  author={Matabuena, Marcos and Petersen, Alexander},
  journal={Journal of the Royal Statistical Society Series C: Applied Statistics},
  volume={72},
  number={2},
  pages={294--313},
  year={2023},
  publisher={Oxford University Press US}
}

@article{ghodrati2022distribution,
  title={Distribution-on-distribution regression via optimal transport maps},
  author={Ghodrati, Laya and Panaretos, Victor M},
  journal={Biometrika},
  volume={109},
  number={4},
  pages={957--974},
  year={2022},
  publisher={Oxford University Press}
}

@article{pegoraro2022projected,
  title={Projected statistical methods for distributional data on the real line with the Wasserstein metric},
  author={Pegoraro, Matteo and Beraha, Mario},
  journal={Journal of Machine Learning Research},
  volume={23},
  number={37},
  pages={1--59},
  year={2022}
}

@article{huang2013oracle,
  title={Oracle inequalities for the lasso in the Cox model},
  author={Huang, Jian and Sun, Tingni and Ying, Zhiliang and Yu, Yi and Zhang, Cun-Hui},
  journal={Annals of statistics},
  volume={41},
  number={3},
  pages={1142},
  year={2013}
}

@article{roth2004generalized,
  title={The generalized LASSO},
  author={Roth, Volker},
  journal={IEEE transactions on neural networks},
  volume={15},
  number={1},
  pages={16--28},
  year={2004},
  publisher={IEEE}
}

@article{mazumder2012graphical,
  title={The graphical lasso: New insights and alternatives},
  author={Mazumder, Rahul and Hastie, Trevor},
  journal={Electronic journal of statistics},
  volume={6},
  pages={2125},
  year={2012}
}

@article{friedman2008sparse,
  title={Sparse inverse covariance estimation with the graphical lasso},
  author={Friedman, Jerome and Hastie, Trevor and Tibshirani, Robert},
  journal={Biostatistics},
  volume={9},
  number={3},
  pages={432--441},
  year={2008},
  publisher={Oxford University Press}
}

@article{verissimo2016degreecox,
  title={DegreeCox--a network-based regularization method for survival analysis},
  author={Ver{\'\i}ssimo, Andr{\'e} and Oliveira, Arlindo Limede and Sagot, Marie-France and Vinga, Susana},
  journal={BMC bioinformatics},
  volume={17},
  number={Suppl 16},
  pages={449},
  year={2016},
  publisher={Springer}
}

@article{sun2014network,
  title={Network-regularized high-dimensional Cox regression for analysis of genomic data},
  author={Sun, Hokeun and Lin, Wei and Feng, Rui and Li, Hongzhe},
  journal={Statistica Sinica},
  volume={24},
  number={3},
  pages={1433},
  year={2014}
}

@article{sennhenn2016structured,
  title={Structured fusion lasso penalized multi-state models},
  author={Sennhenn-Reulen, Holger and Kneib, Thomas},
  journal={Statistics in Medicine},
  volume={35},
  number={25},
  pages={4637--4659},
  year={2016},
  publisher={Wiley Online Library}
}

@inproceedings{zhou2012modeling,
  title={Modeling disease progression via fused sparse group lasso},
  author={Zhou, Jiayu and Liu, Jun and Narayan, Vaibhav A and Ye, Jieping},
  booktitle={Proceedings of the 18th ACM SIGKDD international conference on Knowledge discovery and data mining},
  pages={1095--1103},
  year={2012}
}

@article{cui2021additive,
  title={Additive functional Cox model},
  author={Cui, Erjia and Crainiceanu, Ciprian M and Leroux, Andrew},
  journal={Journal of Computational and Graphical Statistics},
  volume={30},
  number={3},
  pages={780--793},
  year={2021},
  publisher={Taylor \& Francis}
}

@article{qu2016optimal,
  title={Optimal estimation for the functional cox model},
  author={Qu, Simeng and Wang, Jane-Ling and Wang, Xiao},
  journal={The Annals of Statistics},
  pages={1708--1738},
  year={2016},
  publisher={JSTOR}
}

@article{chen2022spatially,
  title={Spatially resolved transcriptomics reveals genes associated with the vulnerability of middle temporal gyrus in Alzheimer’s disease},
  author={Chen, Shuo and Chang, Yuzhou and Li, Liangping and Acosta, Diana and Li, Yang and Guo, Qi and Wang, Cankun and Turkes, Emir and Morrison, Cody and Julian, Dominic and others},
  journal={Acta neuropathologica communications},
  volume={10},
  number={1},
  pages={188},
  year={2022},
  publisher={Springer}
}

@article{zhang2015detection,
  title={Detection of subjects and brain regions related to Alzheimer's disease using 3D MRI scans based on eigenbrain and machine learning},
  author={Zhang, Yudong and Dong, Zhengchao and Phillips, Preetha and Wang, Shuihua and Ji, Genlin and Yang, Jiquan and Yuan, Ti-Fei},
  journal={Frontiers in computational neuroscience},
  volume={9},
  pages={66},
  year={2015},
  publisher={Frontiers Media SA}
}

@article{alzheimer2025,
  title={2025 Alzheimer's disease facts and figures},
  author={{Alzheimer's Association}},
  journal={Alzheimer's \& Dementia},
  volume={21},
  number={4},
  pages={e70235},
  year={2025},
  publisher={Wiley Online Library}
}

@article{gaugler20192019,
  title={2019 Alzheimer's disease facts and figures},
  author={Gaugler, Joseph and James, Bryan and Johnson, Tricia and Marin, Allison and Weuve, Jennifer},
  journal={Alzheimers \& dementia},
  volume={15},
  number={3},
  pages={321--387},
  year={2019},
  publisher={ELSEVIER SCIENCE INC STE 800, 230 PARK AVE, NEW YORK, NY 10169 USA}
}

@article{risacher2009baseline,
  title={Baseline MRI predictors of conversion from MCI to probable AD in the ADNI cohort},
  author={Risacher, Shannon L and Saykin, Andrew J and Wes, John D and Shen, Li and Firpi, Hiram A and McDonald, Brenna C},
  journal={Current Alzheimer Research},
  volume={6},
  number={4},
  pages={347--361},
  year={2009},
  publisher={Bentham Science Publishers}
}

@article{phan2024increased,
  title={Increased cortical thickness in Alzheimer's disease},
  author={Phan, Tony X and Baratono, Sheena and Drew, William and Tetreault, Aaron M and Fox, Michael D and Darby, R Ryan and Alzheimer's Disease Neuroimaging Initiative and Alzheimer's Disease Metabolomics Consortium},
  journal={Annals of neurology},
  volume={95},
  number={5},
  pages={929--940},
  year={2024},
  publisher={Wiley Online Library}
}

@article{liu2010analysis,
  title={Analysis of regional MRI volumes and thicknesses as predictors of conversion from mild cognitive impairment to Alzheimer's disease},
  author={Liu, Yawu and Paajanen, Teemu and Zhang, Yi and Westman, Eric and Wahlund, Lars-Olof and Simmons, Andrew and Tunnard, Catherine and Sobow, Tomasz and Mecocci, Patrizia and Tsolaki, Magda and others},
  journal={Neurobiology of aging},
  volume={31},
  number={8},
  pages={1375--1385},
  year={2010},
  publisher={Elsevier}
}

@article{acosta2012diffusion,
  title={Diffusion tensor metrics as biomarkers in Alzheimer's disease},
  author={Acosta-Cabronero, Julio and Alley, Stephanie and Williams, Guy B and Pengas, George and Nestor, Peter J},
  journal={PloS one},
  volume={7},
  number={11},
  pages={e49072},
  year={2012},
  publisher={Public Library of Science San Francisco, USA}
}

@inproceedings{wang2021deep,
  title={Deep learning based diagnosis of Alzheimer's disease using structural magnetic resonance imaging: A survey},
  author={Wang, Tian and Cao, Lihong},
  booktitle={2021 3rd international conference on applied machine learning (ICAML)},
  pages={408--412},
  year={2021},
  organization={IEEE}
}

\end{document}


\title{Supporting Materials for Penalized Cox Regression with Graph-Structured Distribution-Valued Predictors}

\author{Rahul Ghosal$^{1,\dagger,\ast}$, Dhrubajyoti Ghosh$^{2,\dagger}$, Yuanzhen Yue$^{1}$, and Arkaprava Roy$^{3}$ \\[4pt]
\textit{$^{1}$ Department of Epidemiology and Biostatistics, University of South Carolina \\
$^{2}$ Department of Data Science and Analytics, Kennesaw State University \\
$^{3}$ Department of Biostatistics, University of Florida}
\\[2pt]
{rghosal@mailbox.sc.edu}}

\markboth%
{R. Ghosal and others}
{GFDCM}

\maketitle
\footnotetext{To whom correspondence should be addressed. $^{\dagger}$Co-first authors.}

\section{Web Appendix A: Simulation under a Misspecified Graph (Scenario B)}
\label{sec:scenB}
In Scenario B, the data are generated exactly as in Scenario A of the main paper, but GFDCM is fit with a misspecified graph in which the true edge $(1,2)$ is omitted and the spurious edges $(2,3)$, $(5,6)$, and $(1,8)$ are added. The edges among nodes $\{3,4,5\}$ are retained. Results are based on $M=100$ Monte Carlo replications with $n=300$.

Web Table~\ref{tab:scenB_sel} reports the selection performance. All active nodes were selected in every replicate, while each null node was selected in $12\%$-$15\%$ of the replicates. Web Table~\ref{tab:scenB_fus} reports the fusion performance. None of the spurious edges was fused in any replicate. However, BIC selected little fusion: $\lambda_2=0$ was chosen in $22\%$ of the replicates and $\lambda_2\le 0.17$ in $96\%$, and the number of unique active groups was most often five, i.e., no fusion among the selected active nodes. Web Table~\ref{tab:scenB_mise} reports the MISE of the active distributional effects. Without the variance reduction that fusion provides in Scenario A, the MISE of GFDCM exceeded that of the FCM for all five effects.

One plausible explanation for the limited fusion is that a single fusion parameter $\lambda_2$ governs all edges. Because null nodes are typically not selected ($\hat{\bm\gamma}_6=\hat{\bm\gamma}_8=\mathbf{0}$), the spurious edges $(5,6)$ and $(1,8)$ contribute $\lambda_2 w_{2,56}\|\bm\gamma_5\|_2$ and $\lambda_2 w_{2,18}\|\bm\gamma_1\|_2$ to the penalty, acting as additional sparsity penalties on the active nodes 5 and 1. Values of $\lambda_2$ large enough to fuse nodes $\{3,4,5\}$ therefore also shrink $\beta_1(\cdot)$ and $\beta_5(\cdot)$, and BIC favors small values of $\lambda_2$. Overall, a misspecified graph did not lead to missed active nodes or spurious fusion, but it increased the false positive rate and degraded the estimation performance of GFDCM relative to the correctly specified graph.

\newpage

\section{Web Tables}

\begin{table}[H]
\caption{Descriptive summaries of the variables considered in the ADNI application, stratified by whether they were censored or progressed to AD. The percentages capture the row proportions.}
\label{demo}
\begin{tabular}{lllll}
\hline
\multirow{2}{*}{Characteristic} & \multicolumn{4}{l}{Mean (SD or Proportion)}                                          \\ \cline{2-5} 
                                & Overall (n = 308) & Censored (n = 124) & AD (n = 184) & p-value          \\ \hline
Age                             & 74.9 (7.1)          & 75.3 (7.2)           & 74.6 (7.05)           & 0.37 \\
Gender                          &                     &                      &                      & 0.81 \\
\multicolumn{1}{c}{Male}        & 200 (64.9\%)        & 82 (41\%)           & 118 (59\%)           &                  \\
\multicolumn{1}{c}{Female}      & 108 (35.1\%)        & 42 (38.9\%)           & 66 (61.1\%)           &                  \\

APOE4                  &                     &                      &                      & \textless{}0.001 \\
\multicolumn{1}{c}{0}       & 142 (46.1\%)        & 74 (52.1\%)           & 68 (47.9\%)           &                  \\
\multicolumn{1}{c}{1}      & 129 (41.9\%)        & 40 (31\%)           & 89 (69\%)           &                  \\
\multicolumn{1}{c}{2}     & 37 (12\%)          & 10 (27\%)           & 27 (73\%)           &                  \\ \hline
\end{tabular}
\end{table}

\begin{table}[H]
\centering
\caption{Scenario B (misspecified graph), $n=300$: selection percentages ($\%$) of each node across the Monte Carlo replications, with the average true positive rate (TPR) and false positive rate (FPR) in percentages.}
\label{tab:scenB_sel}
\begin{tabular}{lc}
\toprule
Node & Selection (\%) \\
\midrule
1 & 100 \\
2 & 100 \\
3 & 100 \\
4 & 100 \\
5 & 100 \\
\midrule
6 & 15 \\
7 & 13 \\
8 & 15 \\
9 & 12 \\
10 & 12 \\
\midrule
TPR & 100 \\
FPR & 13.4 \\
\bottomrule
\end{tabular}
\end{table}

\begin{table}[H]
\centering
\caption{Scenario B (misspecified graph), $n=300$: percentages ($\%$) of the number of unique active groups identified across the Monte Carlo replications. CF is the percentage of replicates recovering the best partition attainable under the supplied graph, $\{1\},\{2\},\{3,4,5\}$. The next rows give the percentage of replicates in which each spurious edge was fused, and the last row the percentage of replicates in which BIC selected $\lambda_2=0$.}
\label{tab:scenB_fus}
\begin{tabular}{lc}
\toprule
 & Percentage (\%) \\
\midrule
\multicolumn{2}{l}{Unique groups} \\
\quad 3 & 3 \\
\quad 4 & 3 \\
\quad 5 & 65 \\
\quad 6 & 16 \\
\quad 7 & 8 \\
\quad 8 & 5 \\
\midrule
CF & 3 \\
\midrule
Spurious edge $(2,3)$ fused & 0 \\
Spurious edge $(5,6)$ fused & 0 \\
Spurious edge $(1,8)$ fused & 0 \\
\midrule
$\lambda_2=0$ selected & 22 \\
\bottomrule
\end{tabular}
\end{table}

\begin{table}[H]
\centering
\caption{Scenario B (misspecified graph), $n=300$: mean integrated squared error (MISE) of the active distributional effects $\beta_j(\cdot)$, $j=1,\ldots,5$, from GFDCM and the functional Cox model (FCM).}
\label{tab:scenB_mise}
\begin{tabular}{lccccc}
\toprule
Method & $\beta_1(\cdot)$ & $\beta_2(\cdot)$ & $\beta_3(\cdot)$ & $\beta_4(\cdot)$ & $\beta_5(\cdot)$ \\
\midrule
GFDCM & 1.37 & 1.26 & 1.28 & 1.24 & 1.12 \\
FCM   & 1.03 & 1.07 & 0.43 & 0.43 & 0.50 \\
\bottomrule
\end{tabular}
\end{table}

\section{Web Figures}
\begin{figure}[H]
\begin{center}
\includegraphics[width=.9\linewidth , height=.7\linewidth]{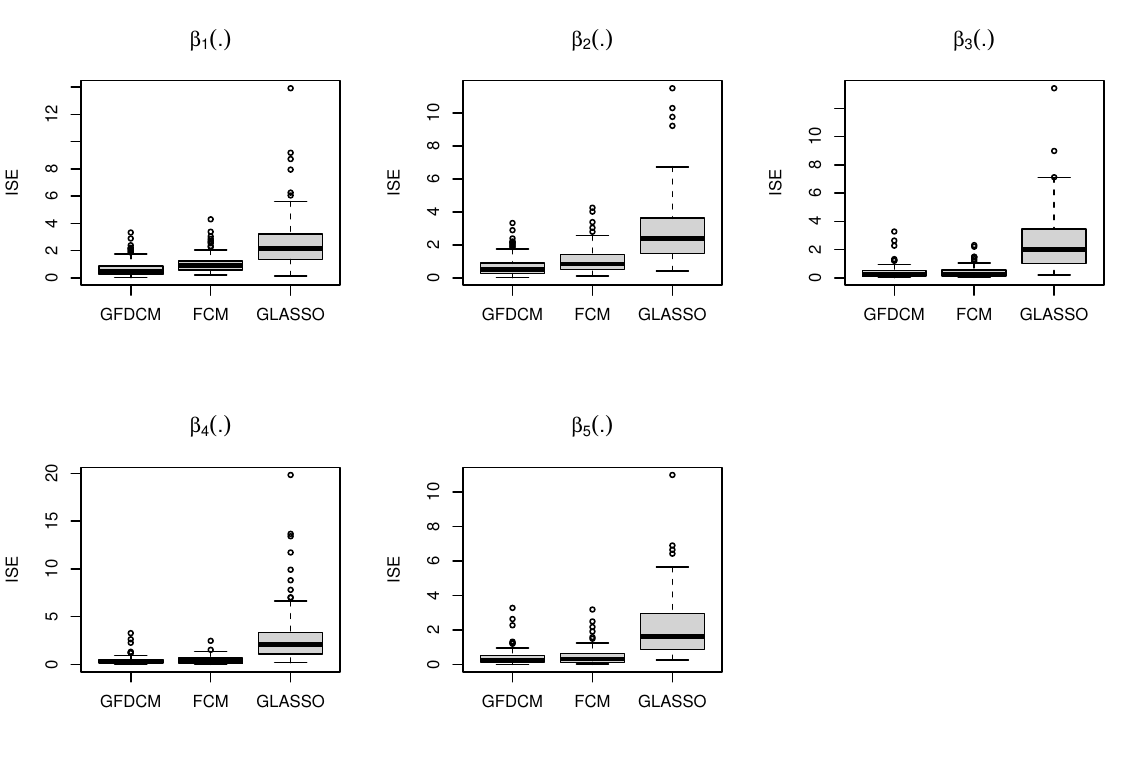}
\caption{The distribution of ISE loss for the three methods; GFDCM, FCM and GLASSO, sample size $n=300$.}
\label{fig:figsim1}
\end{center}
\end{figure}

\begin{figure}[H]
\begin{center}
\includegraphics[width=.9\linewidth , height=.7\linewidth]{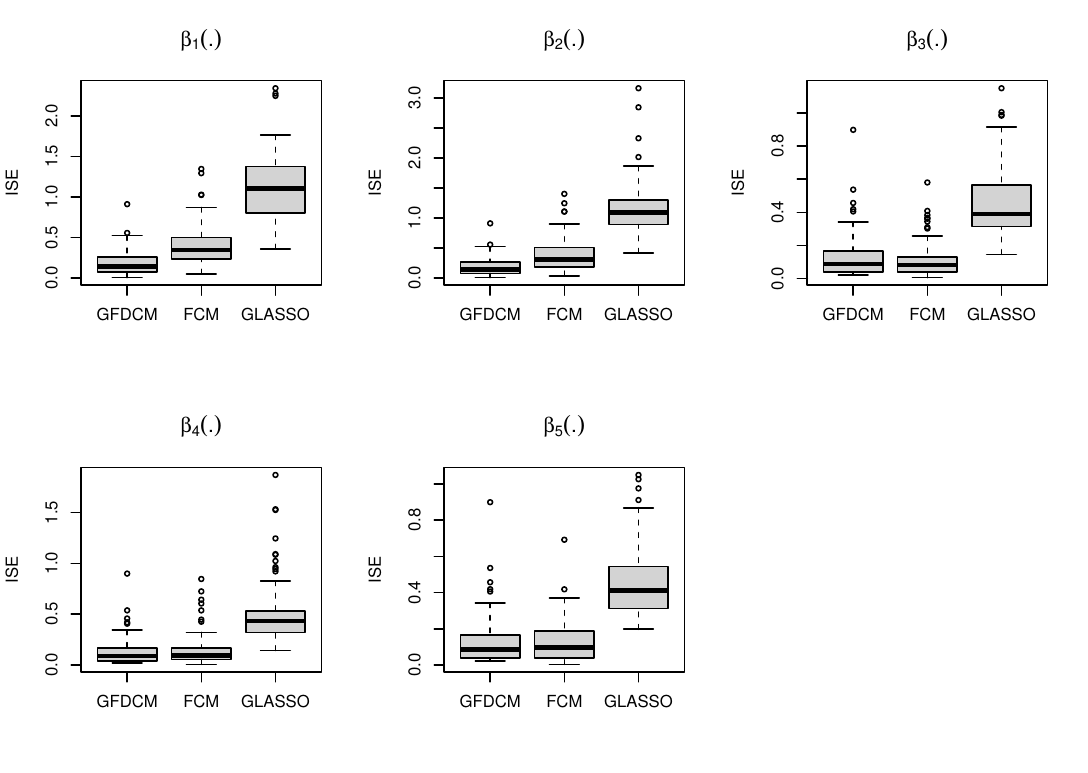}
\caption{The distribution of ISE loss for the three methods; GFDCM, FCM and GLASSO, sample size $n=1000$.}
\label{fig:figsim1}
\end{center}
\end{figure}

\begin{figure}[H]
\begin{center}

\includegraphics[width=.9\linewidth , height=.7\linewidth]{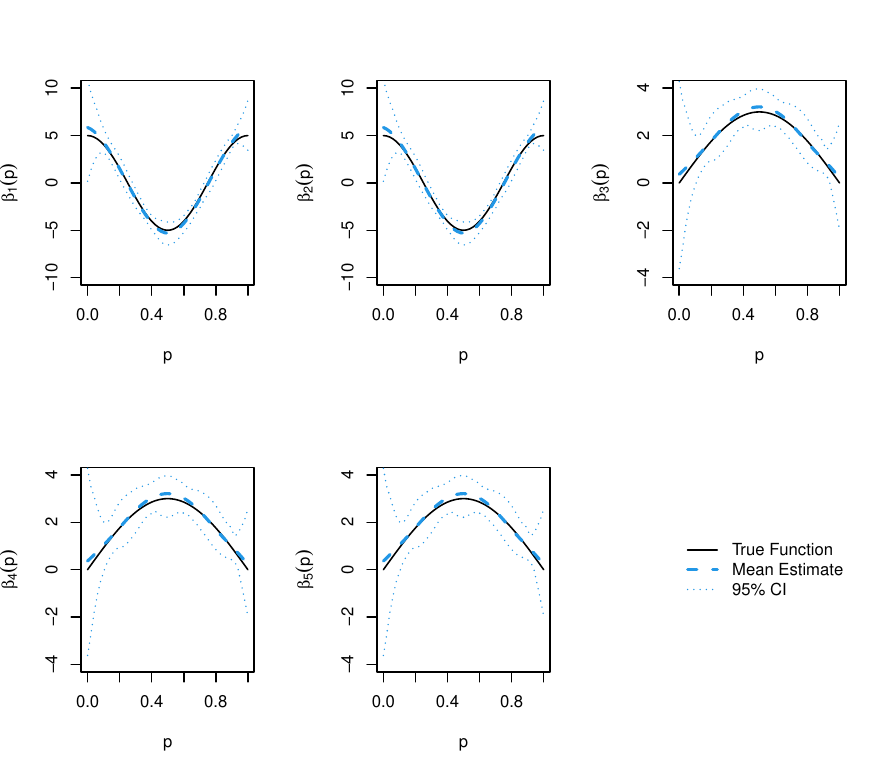}
\end{center}
\caption{Monte Carlo mean estimates from GFDCM and point-wise confidence intervals of the distributional coefficient functions $\beta_j(\cdot)$, for $n=300$.}
\label{fig:fig2s1}
\end{figure}

\begin{figure}[H]
\begin{center}

\includegraphics[width=.9\linewidth , height=.7\linewidth]{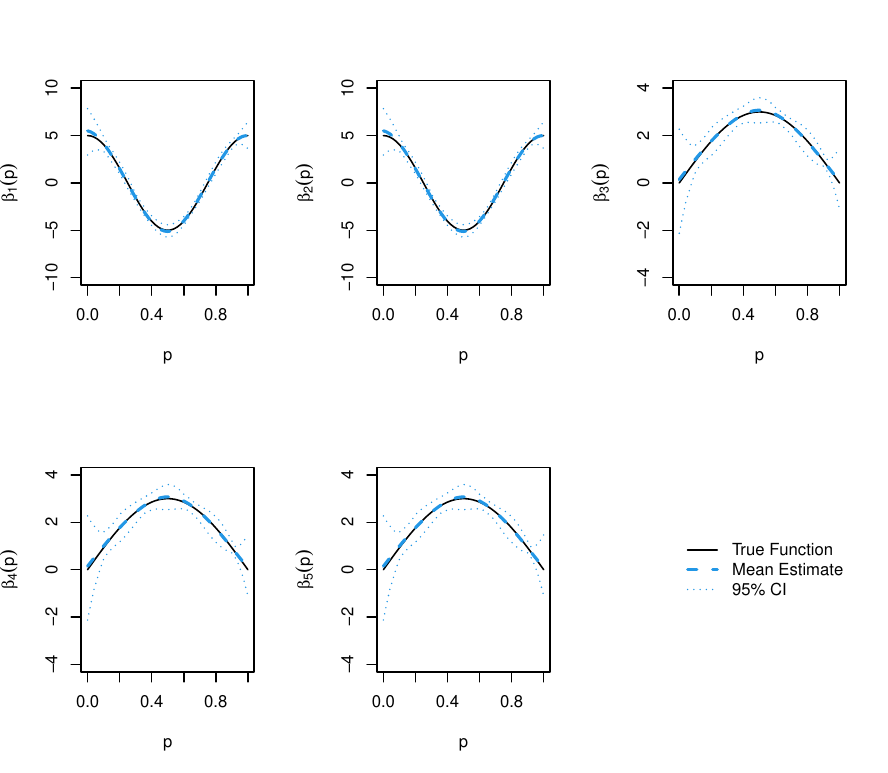}
\end{center}
\caption{Monte Carlo mean estimates from GFDCM and point-wise confidence intervals of the distributional coefficient functions $\beta_j(\cdot)$, for $n=1000$.}
\label{fig:fig2s1}
\end{figure}

\begin{figure}[H]
\begin{center}
\includegraphics[width=.9\linewidth , height=.8\linewidth]{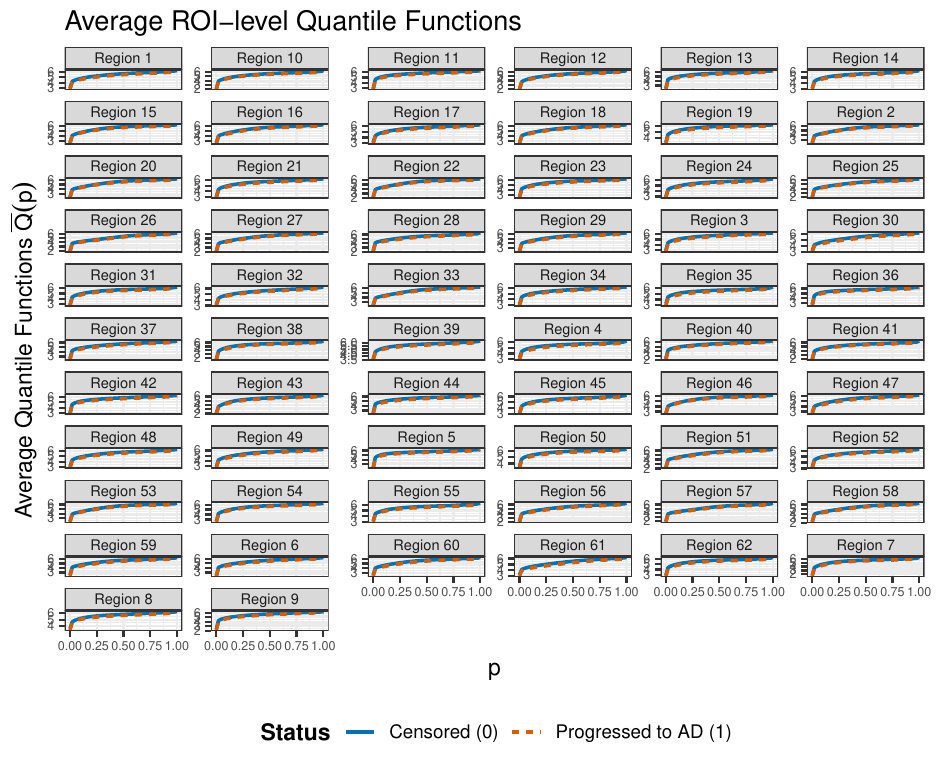}
\end{center}
\caption{Displayed are the averages of ROI-level quantile functions of voxel-intensities (log-transformed) across all 62 regions in the Desikan-Killiany-Tourville (DKT) atlas stratified by the AD status in the ADNI application.}
\label{fig:fig2s1}
\end{figure}

\begin{figure}[H]
\begin{center}
\includegraphics[width=0.6\linewidth , height=0.6\linewidth]{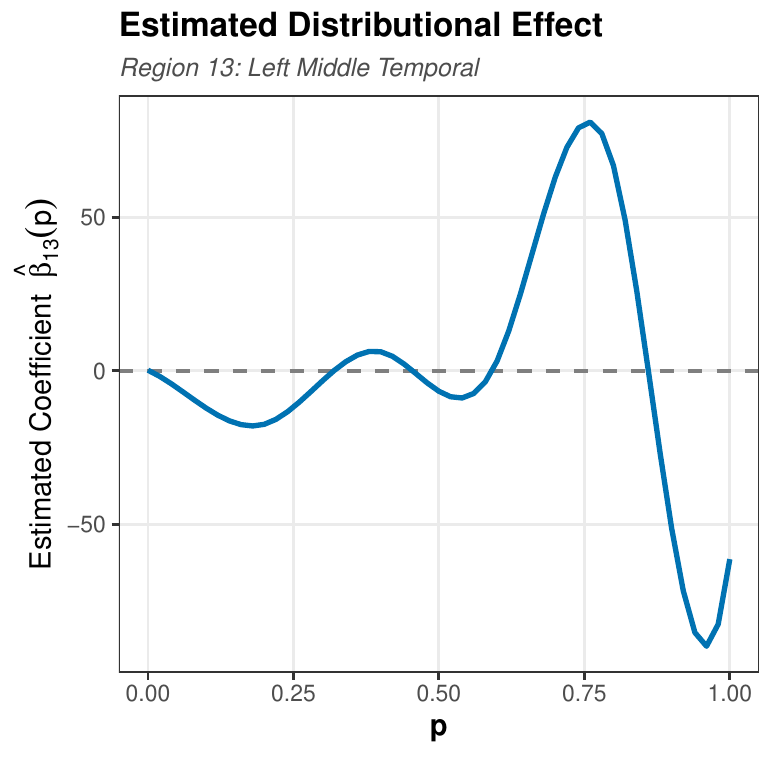}
\end{center}
\caption{The estimated distributional effect $\hat{\beta}_{13}(p)$ of the selected Region 13 (Left Middle Temporal) on log-hazard of progression to AD based on the GFDCM in the ADNI application. This was the second-ranked model in GFDCM according to BIC.}
\label{fig:fig2s1}
\end{figure}

\singlespacing